\documentclass[fleqn,usenatbib]{mnras}

\usepackage{newtxtext,newtxmath}

\usepackage[T1]{fontenc}

\DeclareRobustCommand{\VAN}[3]{#2}
\let\VANthebibliography\thebibliography
\def\thebibliography{\DeclareRobustCommand{\VAN}[3]{##3}\VANthebibliography}

\usepackage{graphicx}	
\usepackage{amsmath}	
\usepackage{placeins}
\usepackage{newtxtext,newtxmath}
\newcommand{\pp}{$\phantom{-}$}
\newcommand{\po}{$\phantom{0}$}
\newbox\grsign \setbox\grsign=\hbox{$>$} \newdimen\grdimen \grdimen=\ht\grsign
\newbox\simlessbox \newbox\simgreatbox
\setbox\simgreatbox=\hbox{\raise.5ex\hbox{$>$}\llap
     {\lower.5ex\hbox{$\sim$}}}\ht1=\grdimen\dp1=0pt
\setbox\simlessbox=\hbox{\raise.5ex\hbox{$<$}\llap
     {\lower.5ex\hbox{$\sim$}}}\ht2=\grdimen\dp2=0pt
\def\simgreat{\mathrel{\copy\simgreatbox}}

\title[Be, V and Cu in the halo star CS~31082-001]{Be, V and Cu in the halo star CS~31082-001 from near-UV spectroscopy}
\author[H. Ernandes et al.]{
H. Ernandes,$^{1}$\thanks{E-mail: heitor.ernandes@usp.br.
Based on observations made with the ESO Very Large
Telescope at Paranal Observatory, Chile; Progr. ID 165.N-0276, and
with the NASA/ESA Hubble Space
Telescope (HST) through the Space Telescope Science Institute, 
operated by the Association of Universities for Research in Astronomy,
Inc., under NASA contract NAS5-26555.}
B. Barbuy,$^{1}$
A. Fria\c ca,$^{1}$
V. Hill,$^{2}$
M. Spite,$^{3}$
F. Spite,$^{3}$
B. V. Castilho$^{4}$
and C. J. Evans$^{5}$ 
\\
$^{1}$Universidade de São Paulo, IAG, Rua do Matão 1226, Cidade Universitária, São Paulo 05508-900, Brazil\\
$^{2}$ Laboratoire Lagrange, Universit\'e de Nice Sophia-Antipolis, Observatoire de la C\^ote d’Azur, Nice 06300 Franc\\
$^{3}$GEPI, Observatoire de Paris, PSL Research University, CNRS,
Place Jules Janssen, 92190 Meudon, France\\
$^{4}$Laborat\'orio Nacional de Astrof\'{\i}sica, Minist\'erio da Ci\^encia,
Tecnologia e Inova\c{c}\~oes - LNA/MCTI, R. Estados Unidos, 154, \\
Itajub\'a, 37504-364, Brazil\\
$^{5}$UK Astronomy Technology Centre, Royal Observatory, Blackford Hill, Edinburgh, EH9 3HJ, UK
}

\date{Accepted 2021 December 22. Received 2021 December 22; in original form 2021 September 27}

\pubyear{2022}

\begin{document}
\label{firstpage}
\pagerange{\pageref{firstpage}--\pageref{lastpage}}
\maketitle
\begin{abstract}
  The `First Stars' programme revealed the metal-poor halo star CS~31082-001 to be r-process and  actinide rich, including the first measurement of a uranium abundance for an old star. To better characterise and understand such rare objects, we present the first abundance estimates of three elements (Be, V, Cu)  for CS~31082-001 from analysis of its near-ultraviolet spectrum. Beryllium is rarely measured in giant stars, and we confirm that its abundance in this star is low due to the rather cool effective temperature that causes destruction of both Be and Li in its atmosphere. Vanadium and copper are iron-peak elements that are starting to be used as chemical-tagging indicators to investigate the origin of stellar populations. We find V and Cu abundances for CS~31082-001 that are comparable to other metal-poor stars, and present new chemical evolution models to investigate our results. In the case of V, extra nucleosynthesis due to interaction of neutrinos with matter is included in the models to be able to reproduce the measured abundance. Given the availability of high-quality spectroscopy of CS~31082-001, we also explore other atomic lines in the near-ultraviolet as a template for future studies of metal-poor stars with the planned CUBES instrument in development for the Very Large Telescope.

\end{abstract}
\begin{keywords}
stars: abundances -- stars: atmospheres -- stars: individual: BPS CS~31082-001 -- Galaxy: halo
\end{keywords}



\section{INTRODUCTION}

Field stars in the Galactic halo are thought to have been formed in an early inner halo or accreted from disrupted dwarf galaxies. Detailed abundance studies (so-called `Galactic archaeology') are a powerful technique to help us understand the origin of these halo populations. To date, the iron-peak elements have been less studied than alpha and neutron-capture elements, but they can be important indicators of both the processes and sources of nucleosynthesis  (e.g. Minelli et al. 2021).

In their seminal work, Cayrel et al. (2004) described the `First Stars' Large Programme at the European Southern Observatory (ESO), that
included 37 nights of observations of metal-poor stars with the Ultraviolet and Visual Echelle Spectrograph (UVES, Dekker et al. 2000) on the Very Large Telescope (VLT). An important target in this programme was the metal-poor halo star CS~31082-001, with [Fe/H]\,$=$\,$-$2.9 and $V$\,$=$\,11.6\,mag. Its abundances were presented in a series of papers, including the first detection of uranium in a stellar spectrum (enabling an age estimate from the ratio of the U abundance to that of thorium, Cayrel et al. 2001), abundance determinations of lighter (Li to Zn) and heavy (Sr to U) elements (Hill et al. 2002), and further study of C to Zn (Cayrel et al. 2004). Ultraviolet (UV) observations with the Space Telescope Imaging Spectrograph (STIS) on the {\em Hubble Space Telescope} ({\em HST}) opened-up further abundance studies of the star, confirming it to be rich in both r-process elements and actinides (Barbuy et al. 2011; Siqueira-Mello et al. 2013).

A subset of metal-poor stars are found to be actinide rich. Holmbeck et al. (2020) recently redefined the r-II sub-class of r-process enhanced stars as those characterized by [Eu/Fe] $\geq$ 0.7, instead of the previous definition of [Eu/Fe] $\geq$ $+$1.0 (Beers \& Christlieb 2005), giving a total of 72 stars now classified as r-II. It has only been possible to measure the actinide uranium abundance in six stars (Holmbeck et al. 2018), namely, CS 31082-001 (Hill et al. 2002), BD +17$^{\circ}$3248 (Cowan et al. 2002),
HE1523-0901 (Frebel et al. 2007), 
RAVE J203843.2-002333 (Placco et al. 2017), 
CS 29497-004 (Hill et al. 2017) and
2MASS J09544277+5246414 (Holmbeck et al. 2018).
Upper limits were reported for others, including CS22892-052 (Sneden et al. 2003) and J1538-1804 (Sakari et al. 2018). CS~31082-001 and J09544277+5246414
are clear actinide-boost stars, showing a large enhancement of the actinides relative to any other class of old stars.
For this reason CS~31082-001 is a reference star in this
respect (e.g. Sneden et al. 2008).


Given the limited efficiency of current facilities, the ground-UV region (300-400\,nm) is less explored in the context of stellar abundances than longer, visible wavelengths. Partly motivated by the development of a new, UV-optimised spectrograph for the VLT, the Cassegrain U-Band Efficient Spectrograph (CUBES), here we revisit the UVES data of CS~31082-001 to investigate its abundances of three elements observed in the near-UV that were not included in previous analyses, namely beryllium, vanadium and copper. 

Be, like Li, is a fragile element. In stellar interiors it is destroyed as soon as the temperature is higher than 3.5\,$\times$\,10$^{6}$K. But, unlike Li, Be is not significantly produced by primordial nucleosynthesis. In very metal-poor stars it is formed through  cosmic-ray spallation, with the latter probably emitted by the first supernovae (Reeves  et al. 1970).
 V and Cu are iron-peak elements, that
are gaining importance in their use as indicators of the early nucleosynthesis processes that enriched old stars.
We also compare abundances estimated from the near-UV for other iron-peak elements with those published from diagnostic lines at longer optical wavelengths.

This paper is structured as follows. The UVES spectroscopy and previous abundances are summarised in Section~2, with our estimated abundances of Be and iron-peak elements from the near-UV analysis reported in Sections~3 and 4, respectively. In Section 5 we discuss chemical evolution models to account for our results, and in Section~6 we briefly summarise our results. 

In addition to providing new constraints on models attempting to reproduce the observed abundance pattern in CS~31082-001, we have used its spectra as a template to investigate future studies of metal-poor stars with CUBES; illustrative results from these tests are presented in the Appendix. 


\section{OBSERVATIONS AND METHODS}

\subsection{Near-UV spectroscopy}
In our new analysis we use the mean combined spectrum of the three 1\,hr exposures obtained by the First Stars programme (ID 165.N-0276, PI: Cayrel) on 2001 October 19-21, using the bluest setting of UVES (with a central wavelength of 346\,nm).  These observations were
obtained with a relatively narrow slit of 0\farcs45, giving $R$\,$\sim$\,77,000 (see Hill et al. 2002), and are
the same reduced data analysed by Siqueira-Mello et al. (2013). 

 The combined UVES data were the primary source spectrum used in our analysis, supported by comparisons with the {\em HST} data. The latter were obtained with the E230M mode of STIS, giving $R$\,$\sim$\,30,000 over the  
1575–3100\,\AA\ range, with a S/N of $\sim$40 from a 45-orbit programme (see Barbuy et al. 2011, Siqueira-Mello et al. 2013).


\subsection{Abundance analysis}

For our abundance analysis we adopted the stellar parameters from Hill et al. (2002), as summarised in Table \ref{cs31}. Model atmospheres were interpolated in the MARCS grids (Gustafsson et al. 2008). For the spectral synthesis calculations we employed the {\tt Turbospectrum} code (Alvarez \& Plez, 1998; Plez, 2012), in which the atomic line lists are essentially those from VALD 
(Ryabchikova et al. 2015) and also including molecular lines of CH, OH, CN and NH.  It is important to note that this code includes scattering in the continuum, that significantly depresses the continuum in the UV region
(e.g. Fig.~1 from Barbuy et al. 2011). This effect needs
  to be taken into account because, when measuring weak lines such
  as the Be~II doublet, the lowered continuum decreases the line depth and
  the abundance can be underestimated.

\begin{table}
\begin{center}
\caption{Stellar parameters adopted for CS~31082-001 (Hill et al. 2002), HD~122563 (Cayrel et al. 2004)}, and the reference star Arcturus (Mel\'endez et al. 2003).
\label{cs31}
\begin{tabular}{lccccccccc}
\hline
\hline
\hbox{Star Name} & \hbox{T$_{\rm eff}$ (K)} & \hbox{log~g} &  \hbox{[Fe/H]} & \hbox{v$_{t}$ (km.s$^{-1}$)} \\
\hline
CS~31082-001 & 4825$\pm$50 & 1.5$\pm$0.3 & $-$2.9$\pm$0.1 & 1.8$\pm$0.2 \\
 HD~122563 & 4600$\pm$50 & 1.1$\pm$0.4 & $-$2.8$\pm$0.1 & 2.0$\pm$0.2 \\ 
Arcturus & 4275$\pm$50 & 1.55$\pm$0.1 & $-$0.54$\pm$0.06 & 1.65 $\pm$0.05 \\
\hline
\end{tabular}
\end{center}
\end{table}

Here we focus on the lines of Be and iron-peak elements in the 300-400\,nm region. Prior to our analysis we compiled the existing results from the First Stars series for light(er) elements (Table~\ref{light}) and neutron-capture elements (Table~\ref{heavy})\footnote{Small differences of 0.01 dex between the [X/Fe] values from Hill et al. (2002), and Cayrel et al. (2004) or Spite et al. (2005) arise from adopting a metallicity of [Fe~I/H]\,$=$
\,$-$2.90 or a mean of [(Fe~I\,$+$\,Fe~II)/H]\,$=$\,$-$2.91.}

 The nitrogen abundance was tentatively deduced from the CN band, although noting that this is very weak in CS~31082-001 and that only an upper limit of A(N)\,$<$\,5.22 was possible (Hill et al. 2002; Cayrel et al. 2004). 
Spite et al. (2005) later estimated A(NH)\,$=$\,4.90 using the molecular data of the NH band from Kurucz linelists. In the same paper it was shown that, in general, there is a systematic difference of $\sim$0.4\,dex between the abundance deduced from the CN and the NH bands. Therefore, Spite et al. (2005) adopted A(N)\,$=$\, 4.50 after correction. We carried out a new measurement of the NH band based on a recent NH linelist (Fernando et al. 2018) and concluded that A(N)\,$=$\,4.57.

In Table~\ref{heavy} we also include abundance estimates from Barbuy et al. (2011) and Siqueira-Mello et al. (2013) obtained from the {\em HST} data alone (i.e. using lines at $<$300\,nm, hence some of the values quoted here are slightly different from the final results from Siqueira-Mello et al.).



\begin{table}
\begin{center}
\caption{Published light-element abundances for CS~31082-001. 
Refs: 1)~Hill et al. (2002); 2) Cayrel et al. (2004); 3) Spite et al. (2005); 4) NLTE calculations from Spite et al. (2011), 5) present work.}\label{light}
\begin{tabular}{lcccl}
\hline
\hline
\hbox{Species} & \hbox{Z} &  \hbox{A(X)} & \hbox{[X/Fe]}\\
\hline
Li~I & 3 & 0.85$^{1,2,3}$ & \ldots \\
C    & 6 & 5.82$^{1,2,3}$    &  +0.2$^{1}$, +0.21$^{2,3}$ \\
N    & 7 & $<$5.22$^{1,2,3}$, 4.90$^{3}$, 4.57$^{5}$ & +0.21$^{1,2,3}$,  $-$0.11$^{3}$,-0.49$^{5}$\\
O~I  & 8 &  6.52$^{1}$, 6.46$^{2}$ & +0.66$^{1}$, +0.6$^{2,3}$\\
Na~I & 11 & 3.70$^{1,2}$ & +0.27$^{1}$, +0.28$^{2}$     \\
Mg~I & 12 &  5.04$^{1,2}$    & +0.36$^{1}$, +0.37$^{2}$    \\
Al~I & 13 &  2.83$^{1,2}$    & $-$0.74$^{1}$, $-$0.73$^{2}$     \\
Si~I & 14 &  4.89$^{1,2}$     & +0.24$^{1}$, +0.25$^{2}$ \\
S   & 16 &  4.54$^{4}$    & +0.36$^{4}$ \\
K~I  & 19 &  2.87$^{1,2}$    & +0.65$^{1}$, +0.66$^{2}$ \\
Ca~I & 20 &  3.87$^{1,2}$    & +0.41$^{1}$, +0.42$^{2}$    \\
Sc~II & 21 & 0.28$^{1,2}$     & +0.01$^{1}$, +0.02$^{2}$ \\
Ti~I & 22 & 2.37$^{1,2}$     & +0.25$^{1}$, +0.26$^{2}$ \\
Ti~II & 22 & 2.43$^{1,2}$     &  +0.31$^{1}$, +0.32$^{2}$  \\
Cr~I & 24 &  2.43$^{1,2}$    & $-$0.34$^{1}$, $-$0.33$^{2}$  \\
Mn~I & 25 &  2.14$^{1}$, 1.98$^{2}$    &  $-$0.35$^{1}$, $-$0.50$^{2}$  \\
Fe~I & 26 &  4.60$^{1,2,3,4}$    &  $\phantom{+}$0.00$^{1,2,3,4}$   \\
Fe~II & 26 &  4.58$^{1,2,3,4}$    & $\phantom{+}$0.00$^{1,2,3,4}$   \\
Co~I & 27 & 2.28$^{1,2}$    & +0.26$^{1}$, +0.27$^{2}$    \\
Ni~I & 28 & 3.37$^{1,2}$    &  +0.02$^{1}$, +0.03$^{2}$   \\
Zn~I & 30 & 1.88$^{1,2}$     &  +0.18$^{1}$, +0.19$^{2}$ \\
\hline
\end{tabular}
\end{center}
\end{table}

\begin{table}
\begin{center}
\caption{Published heavy-element abundances for CS~31082-001. Refs: 1)~Hill et al. (2002); 2) Plez et al. (2004); 3) Barbuy et al. (2011); 4) Siqueira-Mello et al. (2013).}\label{heavy}
\begin{tabular}{lccccl}
\hline
\hline
\hbox{Ele.} & \hbox{Z} &  \hbox{A(X)$_{\rm VLT}$} &
\hbox{A(X)$_{\rm HST}$} & \hbox{[X/Fe]}  \\ 
\hline
Ge & 32 & +0.10$^{4}$ & \ldots & -0.55$^{4}$ \\
Sr & 38 & +0.72$^{1}$ & \ldots & +0.65$^{1}$ \\
Y   & 39 & $-$0.23$^{1}$, $-$0.15$^{4}$ & \ldots &  +0.43$^{1}$, +0.53$^{4}$ \\
Zr   & 40 & +0.43$^{1}$ & +0.55$^{4}$ & $<$+0.73$^{1}$,+0.85$^{4}$ \\ 
Nb  & 41 &  $-$0.55$^{1}$ & $-$0.52$^{4}$ & +0.93$^{1}$, +0.96$^{4}$ \\ 
Mo  & 42 & \ldots   & $-$0.11$^{4}$ & +0.90$^{4}$ \\ 
Ru & 44 & +0.36$^{1}$ &  +0.65$^{4}$ & +1.42$^{1}$, +1.71$^{4}$ \\ 
Rh & 45 &  $-$0.42$^{1,4}$    & \ldots & +1.36$^{1,4}$ \\ 
Pd & 46 &  $-$0.05$^{1}$, $-$0.09$^{4}$    & \ldots& +1.16$^{1}$, +1.12$^{4}$ \\ 
Ag & 47 &  $-$0.81$^{1}$, $-$0.84$^{4}$     &  \ldots &+1.15$^{1}$, +1.12$^{4}$ \\ 
Ba  & 56 & +0.40$^{1}$    & \ldots&+1.17$^{1}$  \\ 
La & 57 &  $-$0.60$^{1}$    & \ldots&+1.13$^{1}$ \\ 
Ce & 58 & $-$0.31$^{1,4}$     &\ldots & +1.01$^{1,4}$ \\ 
Pr & 59 & $-$0.86$^{1}$     & \ldots& +1.33$^{1}$ \\ 
Nd & 60 & $-$0.13$^{1}$, $-$0.21$^{4}$     &  \ldots& +1.27$^{1}$, +1.19$^{4}$ \\ 
Sm  & 62 &  $-$0.51$^{1}$, $-$0.42$^{4}$    &\ldots & +0.00$^{1}$, +0.11$^{4}$ \\ 
Eu & 63 &  $-$0.76$^{1}$ & $-$0.75$^{4}$ &  +1.63$^{1}$, +1.64$^{4}$ \\ 
Gd  & 64 &  $-$0.27$^{1}$ & $-$0.22$^{4}$   & +1.51$^{1}$, +1.46$^{4}$ \\ 
Tb & 65 & $-$1.26$^{1}$ & $-$0.50$^{4}$   & +1.74$^{1}$, +0.98$^{4}$ \\ 
Dy & 66 & $-$0.21$^{1}$, $-$0.12$^{4}$ & \ldots  &  +1.55$^{1}$, +1.46$^{4}$ \\ 
Er & 68 & $-$0.27$^{1}$ & $-$0.20$^{4}$     &  +1.70$^{1}$, +1.63$^{4}$ \\ 
Tm & 69 & $-$1.24$^{1}$, $-$1.18$^{4}$ & \ldots    & +1.66$^{1}$, +1.60$^{4}$ \\ 
Hf & 72    &  $-$0.59$^{1}$, $-$0.73$^{4}$  & \ldots  & +1.43$^{1}$, +1.29$^{4}$    \\ 
Os &  76   &  +0.43$^{1}$  & $-$0.07$^{4}$   & +1.30$^{1}$, +1.72$^{4}$ \\ 
Ir & 77 & +0.20$^{1}$  & +0.18$^{4}$     & +1.75$^{1}$, +1.72$^{4}$ \\ 
Pt & 78 & \ldots  & +0.30$^{3}$ & +1.46$^{3}$ \\ 
Au & 79 & \ldots  & $-$1.00$^{3}$ & +0.89$^{3}$  \\ 
Pb & 82    &  $-$0.55$^{2}$ & $-$0.65$^{3}$    & +0.40$^{2}$, +0.30$^{3}$  \\ 
Bi & 83 & \ldots  & $-$0.40$^{3}$ & +1.83$^{3}$ \\ 
Th & 90 & $-$0.98$^{1}$    & \ldots     &  +1.83$^{1}$ \\
U &  92 & $-$1.92$^{1}$ &  \ldots     &  +1.49$^{1}$ \\
\hline
\end{tabular}
\end{center}
\end{table}

\section{Beryllium}

\begin{figure}
    \centering
    \includegraphics[width=3.3in]{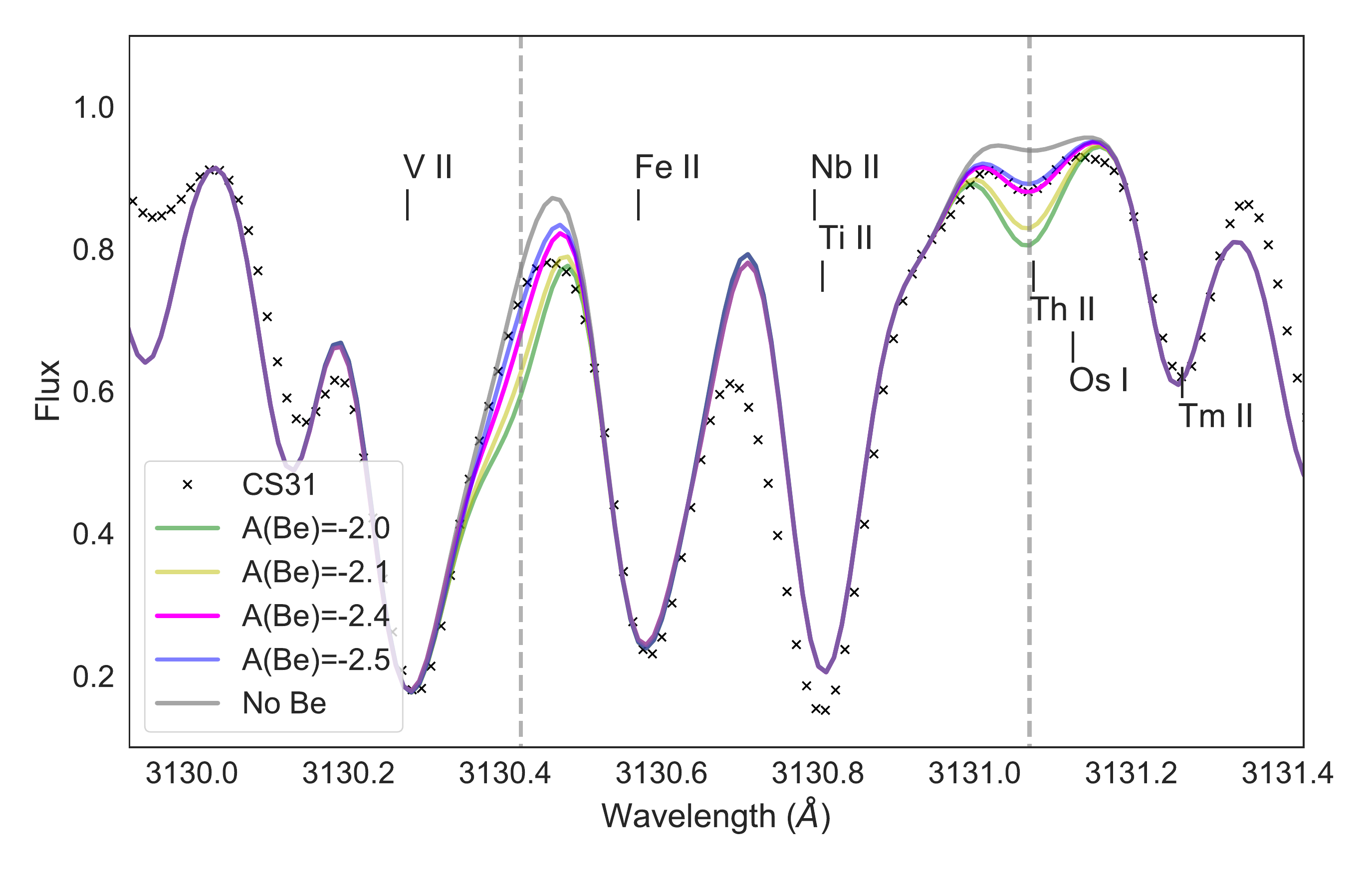}
    \caption{Fits to the Be~II 3130.42 and 3131.07\,\AA\ doublet. Blue, magenta, yellow and green model spectra correspond to A(Be)\,$=$\,$-$2.5,  $-$2.4, $-$2.1 and $-$2.0, respectively. The grey line is the model spectrum without Be.}
    \label{FitBe2}
\end{figure}

The derivation of Be abundances for metal-poor stars is challenging because the Be~II doublet (see Table~\ref{loggf} for details) is weak and blended with other nearby absorption lines.
Lines of Mn, Cr, Ni, Ge, Ta, Gd, Tm, Mo are all present in the region but they do not contribute significantly to the blends. In contrast, weak lines of Fe, V, Ti, Th, Os and molecular CNO lines (mainly from OH) do contribute as blends to the Be lines.
In particular, the longer-wavelength Be\,II line is blended with Th~II 3131.07\,\AA\ and Os~I 3131.12\,\AA. Fig.~\ref{FitBe2} shows the spectral synthesis calculation without including Be and with  A(Be) = $-$2.5,  $-$2.4, $-$2.1, and $-$2.0,
suggesting abundances of A(Be) = $-$2.5 and $-$2.4 for CS~31082-001 from the Be~II 3130.42 and 3131.07 {\rm \AA} lines, respectively. 

The Be abundance measured in CS~31082-001 does not, however, indicate its primordial Be abundance. Be is a fragile element destroyed as soon as the temperature reaches 3.5\,$\times$\,10$^6$ K. Nonetheless, Be is less fragile than Li, which is destroyed at 2.5\,$\times$\,10$^6$ K. All the stars in the Spite's Li plateau (Spite \& Spite 1982) have effective temperatures higher than 5900~K. At lower temperatures the convective zone is deeper and, as a consequence, Li is brought to deep layers where it is gradually destroyed, and the same process affects also Be. At the metallicity of CS~31082-001, following e.g. Boesgaard et al. (2011), its original abundance should be about A(Be)\,$=$\,$-$2.2. Destruction of Be in CS~31082-001 is thus confirmed, as might be expected by its low Li abundance (A(Li)\,$=$\,0.85), which is much below the Spite plateau at A(Li)\,$\sim$\,+2.2.

During their evolution, metal-poor giants undergo a first mixing (1st dredge-up) when they leave the subgiant branch. Later, at the level of the bump, they undergo extra mixing: C is transformed into N and Li practically disappears. It has been shown that CS~31082-001 has not yet undergone this extra mixing (Spite et al. 2005), therefore some Li and Be appear to be still present. In contrast, HD~122563 was another giant star observed by the First Stars programme but located after the bump (see stellar parameters in Table~\ref{cs31}). In this star Li is not measurable (A(Li)\,$<$\,0.6,  Spite et al. 2005), and we verified that the Be line appears also to be absent.


\begin{figure}
    \centering
    \includegraphics[width=3.3in]{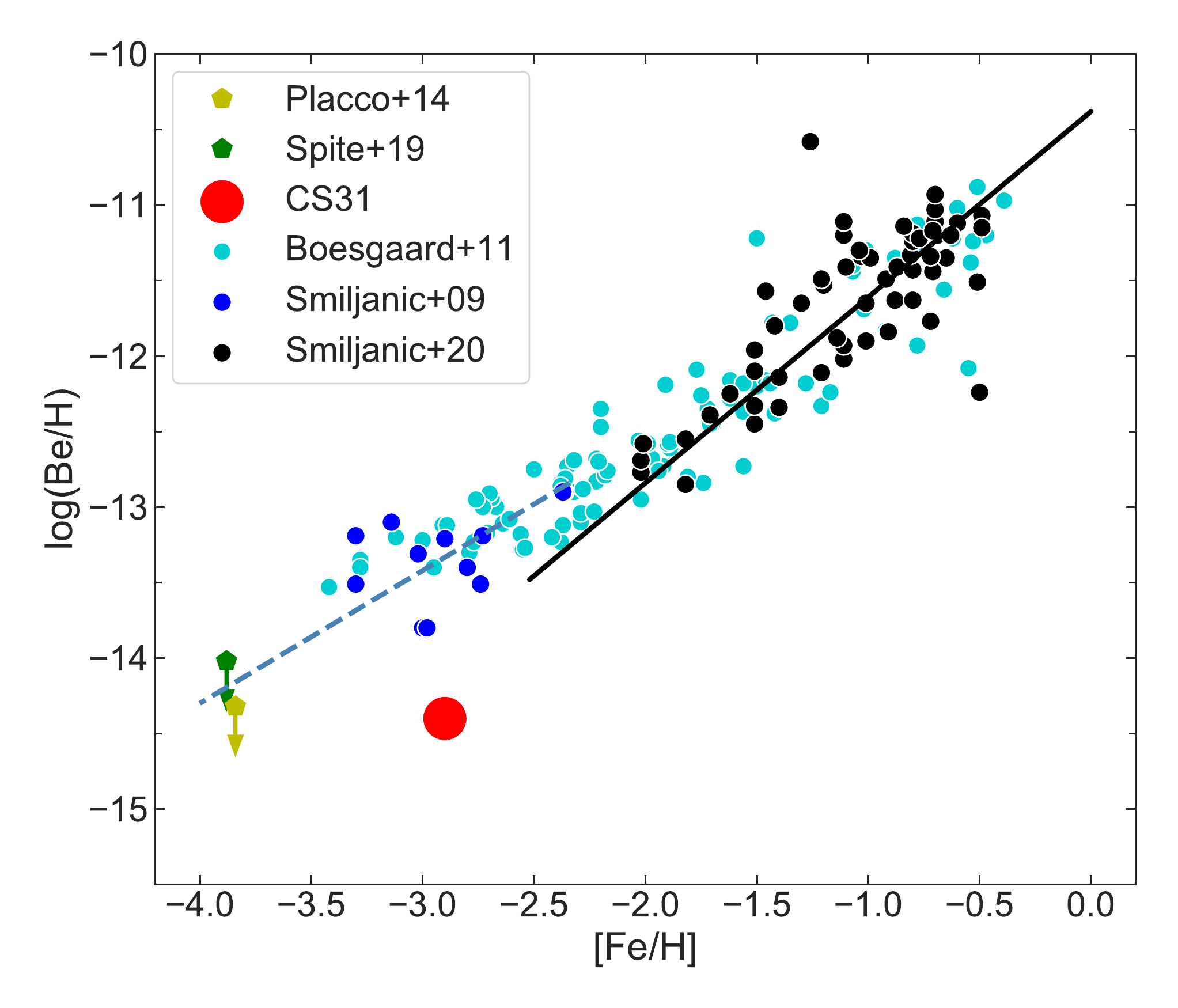}
    \caption{ log(Be/H) vs. [Fe/H] for CS 31082-001 (red circle) compared with literature data as follows: solid blue circles, Smiljanic et al. (2021); solid black circles, Smiljanic et al. (2009); solid green pentagons, Spite et al. (2019); solid yellow pentagons, Placco et al. (2014); solid cyan circles, Boesgaard et al. (2011). The solid line is a fit to the data by Smiljanic et al. (2009). The dashed line is the locus of dwarf metal-poor stars (Spite et al. 2019).}
    \label{Beplot}
\end{figure}

The estimated Be abundance in CS~31082-001 is the first measurement of a Be abundance in a metal-poor giant. Results in the literature are for stars with effective temperatures greater than $\sim$5500 K, and with gravities
around the turn-off. In Fig.~\ref{Beplot} we show literature values for dwarf stars from Boesgaard et al. (2011), Placco et al. (2014), Smiljanic et al. (2009, 2021) and Spite et al. (2019), where it can be seen that the Be abundance increases with [Fe/H]. In CS~31082-001 with [Fe/H]\,$=$\,$-$2.9 we would expect $\rm log(Be/H) \approx -13.8$ for a dwarf star, to be compared with our estimate of log(Be/H)\,$=$\,$-$14.5 for CS~31082-001. In fact, Be (like Li) has been partially destroyed during the first dredge up.  This destruction depends mainly on the extension of the convective zone. The combination of the Be and Li depletion in metal-poor giants observed after the first dredge-up, could bring important constraints on the maximum depth of the convective layer.

\section{Iron-peak elements}

Table \ref{loggf} gives the list of near-UV lines of iron-peak elements considered here, together with oscillator strengths from Kurucz (1993)\footnote{http://kurucz.harvard.edu/atoms.html}, NIST\footnote{https://physics.nist.gov/PhysRefData/ASD/lines\_form.html}, 
VALD\footnote{http://vald.astro.uu.se/}
(Piskunov et al. 1995, Ryabchikova et al. 2015),
and literature values. A more extensive list of the most prominent lines in the near-UV can be found in Ernandes et al. (2020). 

We included lines of the elements from Sc to Ge. Note that Germanium is considered the last of the iron-peak elements (Woosley \& Weaver, 1995, hereafter, WW95), but it is sometimes considered as part of the heavy neutron-capture elements, or else as a trans-iron element (e.g. Peterson et al. 2020). 

\begin{table*}
\caption{ List of near-UV absorption lines considered. Refs: 1) Wood et al. (2014); 2) Den Hartog et al. (2011); 3) Li et al. (1999); 4) Lawler et al. (2013); 5) VALD3; 6) Ou et al. (2020). Adopted log~gf values are indicated in bold face. }   
\label{loggf} 
\centering                  
\begin{tabular}{lccccccccc} 
\hline\hline             
  Ion   &  Wavelength ({\rm \AA}) & $\chi_{\rm ex}$ & log~gf$_{\rm Kur}$ &log~gf$_{\rm VALD}$ & log~gf$_{\rm NIST}$ & log~gf$_{\rm lit}$ & [X/Fe] & $\Delta$[X/Fe]$_{\rm NLTE}$ & Ref. \\
\hline
Be\,II & 3130.42 &  \pp0.00 & $-$0.168  & $-$0.168 & {\bf$-$0.178}  & \ldots & $-$0.98  & \ldots & 5 \\
Be\,II & 3131.07 &  \pp0.00 & $-$0.468  & $-$0.468 & {\bf$-$0.479}  & \ldots & $-$0.98  & \ldots & 5 \\
\hline  
Sc\,II$^{*}$ & 3576.34 & \pp0.0084 & \pp0.130  & \pp0.007  & {\bf \pp0.01} & \ldots & +0.03 & \ldots & 5\\
Sc\,II$^{*}$ & 3590.47 & \pp0.022\  & $-$0.500 & $-$0.552 & {\bf $-$0.55} & \ldots & +0.52  & \ldots& 5\\
\hline  
Ti\,I  & 3199.914 & \pp0.048 &  \pp0.200 & \pp0.310 &  {\bf \pp0.20}  & ~+0.31 & +0.35 & \ldots & 4 \\
Ti\,I  & 3717.391 & \pp0.000 & $-$1.210 & $-$1.228 & {\bf$-$1.20}  & $-$1.19 &  +0.35 & \ldots & 4 \\
Ti\,I  & 3729.807 & \pp0.000 & $-$0.340 & $-$0.280 & {\bf$-$0.289} & $-$0.28 &  +0.35 & \ldots & 4 \\
Ti\,I  & 3924.526 & \pp0.021 & $-$0.937 & $-$0.870 & {\bf$-$0.883} & $-$0.87 &  +0.35 & \ldots & 4 \\
Ti\,I  & 3962.851 & \pp0.000 & $-$1.167 & $-$1.232 &{\bf$-$1.110} & $-$1.10 &   +0.35 & \ldots & 4 \\
Ti\,I$^{*}$  & 3998.64 & \pp0.048\ & $-$0.056 & \pp0.02  & {\bf\pp0.016} & \ldots & +0.02 & \ldots & 4\\
Ti\,II$^{*}$ & 3321.70 & \pp1.231\ & $-$0.320 & $-$0.340 & {\bf$-$0.313} & \ldots & +0.35 & \ldots & 4\\
Ti\,II$^{*}$ & 3343.76 & \pp0.151\ & $-$1.270 & $-$1.180 & {\bf$-$1.149} & \ldots & +0.35 & \ldots & 4\\ 
Ti\,II$^{*}$ & 3491.05 & \pp0.113\ & $-$1.130 & $-$1.100 & {\bf$-$1.153} & \ldots & +0.35 & \ldots & 4\\ 
\hline  
V\,II  & 3517.299 & \pp1.128 & $-$0.310 & \ldots & {\bf$-$0.24} & $-$0.24 & +0.22 & \ldots  & 6 \\
V\,II  & 3545.196 & \pp1.096 & $-$0.390 & \ldots & {\bf$-$0.32} & $-$0.32 & +0.22 & \ldots  & 6 \\
V\,II  & 3715.464 & \pp1.575 & $-$0.380 & \ldots & {\bf$-$0.22} & $-$0.22 & +0.22 & \ldots  & 6 \\
V\,II$^{*}$  & 3951.96 & \pp1.4764 & $-$0.740 & $-$0.730 & {\bf$-$0.73} & $-$0.73 & +0.22 & \ldots & 1 \\
\hline  
Cr\,I$^{*}$  & 3578.68 & \pp0.00 & \pp0.409 & \pp0.42 & {\bf\pp0.408} & \ldots& $-$0.31 & +0.678 & 5\\ 
\hline  
Mn\,II$^{*}$ & 3441.99 & \pp1.776 & $-$0.270 & $-$0.332 & {\bf$-$0.346} & $-$0.346  & $-$0.39 & \ldots & 2 \\
Mn\,II$^{*}$ & 3460.32 & \pp1.809 & $-$0.615 & $-$0.615 & {\bf$-$0.632} & $-$0.631  & $-$0.39 & \ldots & 2 \\
Mn\,II$^{*}$ & 3482.90 & \pp1.833 & $-$0.740 & $-$0.826 & {\bf$-$0.837} & $-$0.837  & $-$0.39 & \ldots & 2 \\
Mn\,II$^{*}$ & 3488.68 & \pp1.847 & $-$0.860 & $-$0.921 & {\bf$-$0.937} & $-$0.937  & $-$0.39 & \ldots & 2 \\
Mn\,II$^{*}$ & 3495.83 & \pp1.855 & $-$1.200 & $-$1.257 & {\bf$-$1.282} & $-$1.280  & $-$0.39 & \ldots & 2 \\
Mn\,II$^{*}$ & 3497.53 & \pp1.847 & $-$1.330 & $-$1.397 & {\bf$-$1.414} & $-$1.418  & $-$0.39 & \ldots & 2 \\
\hline  
Co\,I$^{*}$  & 3412.34 & \pp0.5136 & \pp0.030 & \pp0.030  & {\bf\pp0.03} & \ldots & +0.19 & \ldots & 5  \\
Co\,I$^{*}$  & 3412.63 & \pp0.00   & $-$0.780 & $-$0.780  & {\bf$-$0.78} & \ldots & +0.19 & \ldots & 5  \\
Co\,I$^{*}$  & 3449.16 & \pp0.5815 & $-$0.090 & $-$0.090  & {\bf$-$0.09} & \ldots & +0.19 & \ldots & 5  \\
Co\,I$^{*}$  & 3529.03 & \pp0.1744 & $-$0.880 & $-$0.880  & {\bf$-$0.88} & \ldots & +0.19 & \ldots & 5  \\ 
Co\,I$^{*}$  & 3842.05 & \pp0.9227 & $-$0.770 & $-$0.770  & {\bf$-$0.76} & \ldots & +0.19 & \ldots & 5  \\
Co\,I$^{*}$  & 3845.47 & \pp0.9227 & \pp0.010 & \pp0.010  & {\bf\pp0.01} & \ldots & +0.19 & +0.749 & 5  \\
\hline  
Ni\,I$^{*}$  & 3437.28 & \pp0.00   & $-$1.150 & $-$1.20\po  & {\bf$-$1.15} & \ldots & +0.05 & \ldots & 5 \\ 
Ni\,I$^{*}$  & 3483.77 & \pp0.2748 & $-$1.110 & $-$1.110 & {\bf$-$1.12} & \ldots & +0.05 & \ldots & 5 \\
Ni\,I$^{*}$  & 3500.85 & \pp0.1652 & $-$1.370 & $-$1.270 & {\bf$-$1.37} & \ldots & +0.05 & \ldots & 5\\
Ni\,I$^{*}$  & 3597.71 & \pp0.2124 & $-$1.090 & $-$1.10\po& {\bf$-$1.09} & \ldots & +0.05 & \ldots & 5 \\
Ni\,I$^{*}$  & 3807.14 & \pp0.4228 & $-$1.180 & $-$1.230 & {\bf$-$1.18} & \ldots & +0.05 &\ldots & 5\\
Ni\,I$^{*}$  & 3807.14 & \pp3.8983 & \ldots   & $-$2.816 & {\bf$-$1.18} & \ldots & \ldots & \ldots& Blend \\
\hline  
Cu\,I$^{*}$  & 3247.53 & \pp0.00 & $-$0.062 & $-$0.008 & {\bf$-$0.054} & \ldots & $-$0.79 & \ldots & 5 \\
Cu\,I$^{*}$  & 3273.95 & \pp0.00 & $-$0.359 & $-$0.311 & {\bf$-$0.354} & \ldots & $-$0.79 & \ldots & 5\\
\hline  
Zn\,I$^{*}$  & 3075.90 & \pp0.00   & $-$3.900 & $-$3.900 & {\bf$-$3.80} & \ldots & \ldots & \ldots & 5\\ 
Zn\,I$^{*}$  & 3302.58 & \pp4.0297 & $-$0.057 & $-$0.057 & {\bf$-$0.01} & \ldots & +0.22 & \ldots & 5\\
Zn\,I$^{*}$  & 3345.01 & \pp4.0778 & \pp0.246 & \pp0.246 & {\bf\pp0.30} & \ldots & +0.22 & \ldots & 5 \\ 
\hline  
Ge\,I$^{*}$  & 3039.07 & \pp0.8834 & \ldots  & \pp0.49 & {\bf\pp0.07}& $-$0.08 & \ldots & \ldots & 3 \\ 

\hline
\hline                          
\end{tabular}
\\
\footnotesize{1) Symbols: $^{*}$ Near-UV iron-peak element lines considered by Ernandes et al. 2020;\\
2) The Ge~I line is included for completeness, where its abundance was derived by Siqueira-Mello et al. (2013).}
\end{table*}


     

\subsection{Calculations}

\subsubsection{Hyperfine structure}
To accurately model the near-UV lines of Sc~II, V~II, Mn~II, Co~I, and Cu~I  we investigated their hyperfine structure (HFS) using the code from McWilliam et al. (2013). 
The magnetic dipole A-factor and electronic quadrupole B-factor hyperfine constants used to compute the HFS 
for each species are given in Table~\ref{lines1} with details as follows:

\begin{itemize}
    \item{Scandium: $^{45}$Sc is the only stable nuclide, with nuclear spin I\,$=$\,7/2. Hyperfine constants are from Villemoes et al. (1992) and Kurucz (1993).}\smallskip
    \item{Vanadium: abundances correspond to 99.75\% of $^{51}$V and only 0.25\% of $^{50}$V (Asplund et al. 2009). We therefore adopted $^{51}$V as a unique isotope, with nuclear spin I\,$=$\,7/2. Hyperfine constants are from Wood et al. (2014). Following Ou et al. (2020), HFS splitting was not applied to the V\,II 3715.46 {\rm \AA} line.
    }\smallskip
    \item{Manganese: $^{55}$Mn is the only nuclide that contributes to the abundance (Woodgate \& Martin 1957), with nuclear spin I\,$=$\,5/2. Hyperfine constants are from Den Hartog et al. (2011).}\smallskip
    \item{Cobalt: $^{59}$Co is the only nuclide (Asplund et al. 2009), with nuclear spin I\,$=$\,7/2. Hyperfine constants are from Pickering (1996).}\smallskip
    \item{Copper: Isotopic fractions for Cu~I are 0.6894 for $^{63}$Cu and 0.3106 for $^{65}$Cu (Asplund et al. 2009). The nuclear spin is I\,$=$\,3/2. Abundances were derived from the Cu~I 3247.53 and 3273.95\,\AA\ lines. Hyperfine constants are from Biehl (1976) and Kurucz (1993).}
\end{itemize}


After calculating the line splits and intensities of the HFS lines, we computed synthetic lines 
for the reference star Arcturus, adopting stellar parameters from Mel\'endez et al. (2003), and then comparing them with the observed lines in a near-UV UVES spectrum from the ESO archive\footnote{http://archive.eso.org/dataset/ADP.2020-08-04T15:12:16.253}.  Figures~\ref{ArcCo} and \ref{ArcCu} show the fits to Co and Cu for the Arcturus spectrum. For Sc~II and Mn~II the calculations including the HFS are a poorer match to the observations than those without, possibly because some atomic constants cannot be accounted for; we therefore did not include HFS for these two elements in our analysis of CS~31082-001. We add that for many of the lines studied here, the surrounding lines in the Arcturus spectrum are saturated. The adopted HFS components and corresponding oscillator strengths for the Co~I and Cu~I lines are reported in Tables \ref{hfsCo3}, \ref{hfsCo2} and \ref{hfsCu2}. 

\subsubsection{Non-LTE abundance corrections}\label{NLTE}

As demonstrated by Bergemann \& Cescutti (2010) and Bergemann et al. (2010), NLTE corrections are needed for the Cr and Co lines, with corrections available for two of our diagnostic lines\footnote{http://nlte.mpia.de/gui-siuAC\_secE.php}. For the adopted stellar parameters of CS~31082-001, the NLTE corrections for the Cr~I 3578.68\,{\rm \AA} and Co~I 3845.47\,{\rm \AA} lines are +0.678\,dex and +0.749\,dex, respectively.
In particular, Bergemann \& Cescutti (2010) analysed the discrepancies between abundances derived from Cr~I and Cr~II lines, showing that deficiencies of Cr in metal-poor stars are related to only NLTE effects, and that the Cr/Fe ratio is essentially solar.

\subsection{Abundances: Sc, Ti, Cr, Mn, Co, Ni and Zn}
\begin{figure}
    \centering
    \includegraphics[width=3.0in]{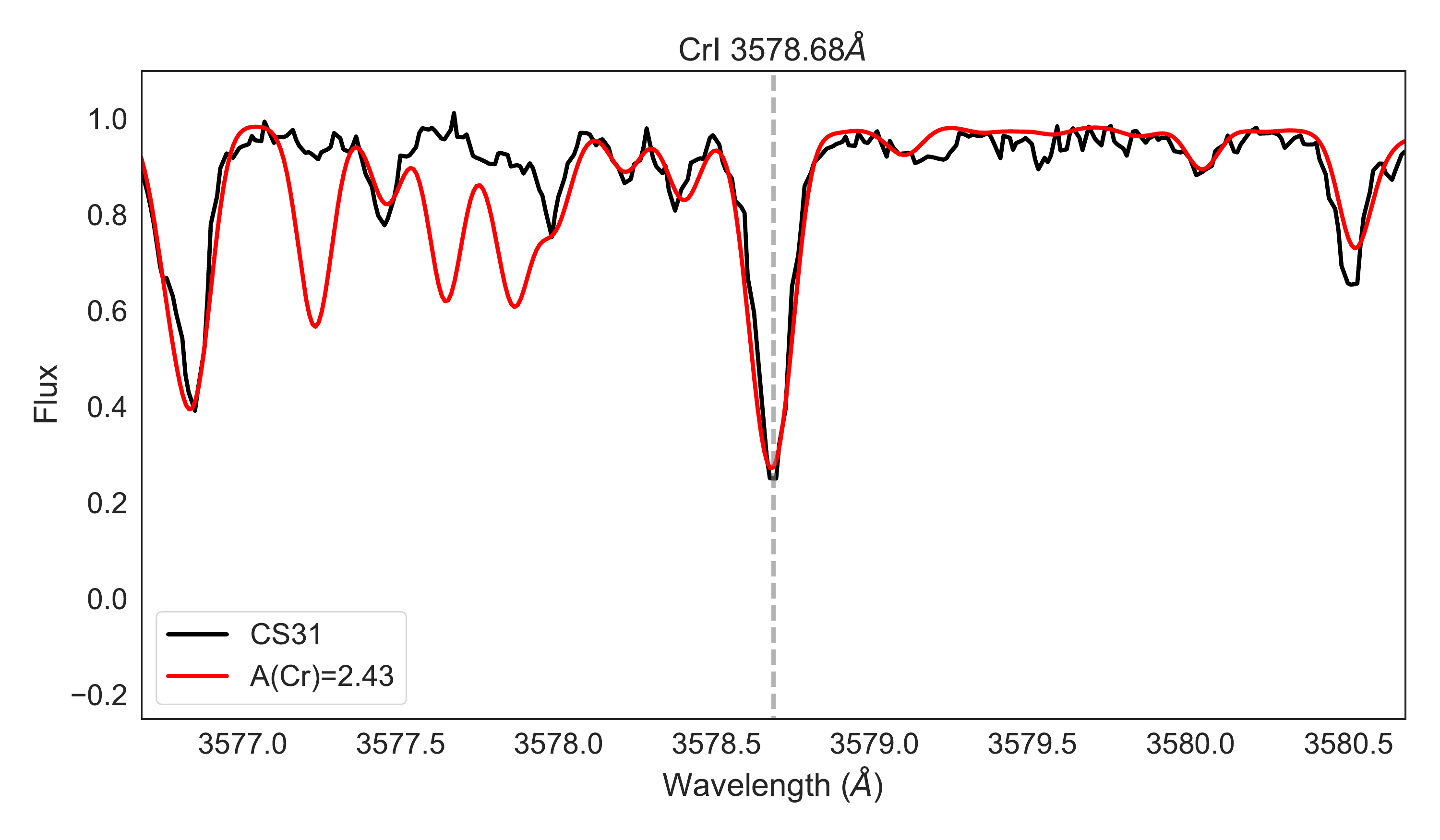}
    \caption{ Best-fit abundance for the Cr~I 3578.68 {\rm \AA} line, computed with [Cr/Fe]\,$=$\,$-$0.31, without NLTE correction (see Sect.~\ref{NLTE}).}
    \label{Cr}
\end{figure}

Adopting the published stellar parameters of CS~31082-001 and varying the abundances of each species to fit the relevant diagnostic lines, our fits to Sc, Ti, Mn, Co, Ni and Zn are shown in Figs.~\ref{Sc} to \ref{Zn}, with the fit to the Cr~I 3578.68\,\AA\ line shown in Fig.~\ref{Cr}.

The fits to the Cr~I, Mn~II, Co~I and Ni~I lines are excellent (Figs.~\ref{Cr}, \ref{Mn}, \ref{Co}, and \ref{Ni}), while those to Sc~II and Ti~I (Figs.~\ref{Sc} and \ref{TiI}, respectively) both have problems. For Sc~II, the two diagnostic lines lead to quite different abundances ($\Delta$A(Sc)\,$=$\,0.49); the Sc~II 3590.47\,\AA\ line appears to be a saturated blend that only has a small dependence on the abundance, so we adopt the value from the Sc~II 3576.34\,\AA\ line.

A similar discrepancy also arises between the Ti~I 3998.64\,\AA\ and 3924.526\,\AA\ lines (see Fig.~\ref{TiI}), where there are blends present in the redward wing. However, the estimated abundance from the Ti~I 3924.526\,\AA\ line also matched that from the four other Ti~I lines in Table~\ref{loggf}. These other lines were included from the line list of Lawler et al. (2013); 
additional Ti~I lines at 3725.1 and 3926.3\,\AA\ were not strong enough in CS~31082-001 to estimate the abundance. Fig \ref{TiI} shows the fits to the Ti~I lines, giving an abundance of A(Ti)\,$=$\,2.40. This is similar to the value obtained from the Ti~II lines (Fig.~\ref{Ti}), so we adopt this value. Compared to the results from optical lines, this result is higher by 0.15\,dex for Ti~I and 0.05\,dex for Ti~II, but still in reasonable enough agreement given the uncertainties.

A summary of the present abundance results with previous values is given in Table~\ref{comparison}. The values for each species are in general agreement within errors.

\subsection{Abundances: V and Cu}
\begin{figure}
    \centering
    \includegraphics[width=3.3in]{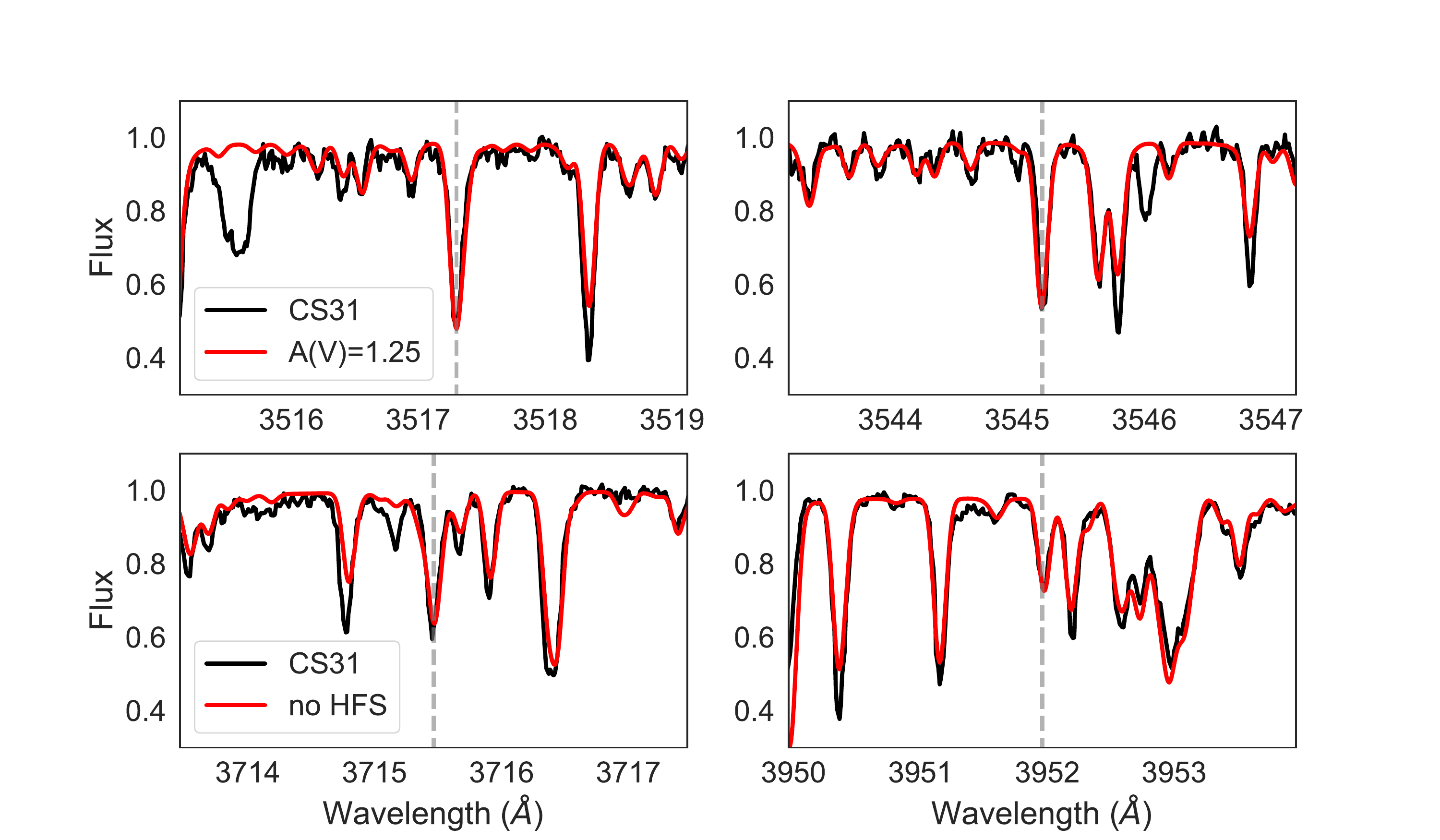}
    \caption{  Best fit abundance for V~II 3517.299, 3545.196, 3715.464 and 3951.96 {\rm \AA} lines, computed with [V/Fe]\,$=$\,+0.22. HFS was not applied to the V~II 3715.464 {\rm \AA} line. }
    \label{V}
\end{figure}

\begin{figure}
    \centering
    \includegraphics[width=3.3in]{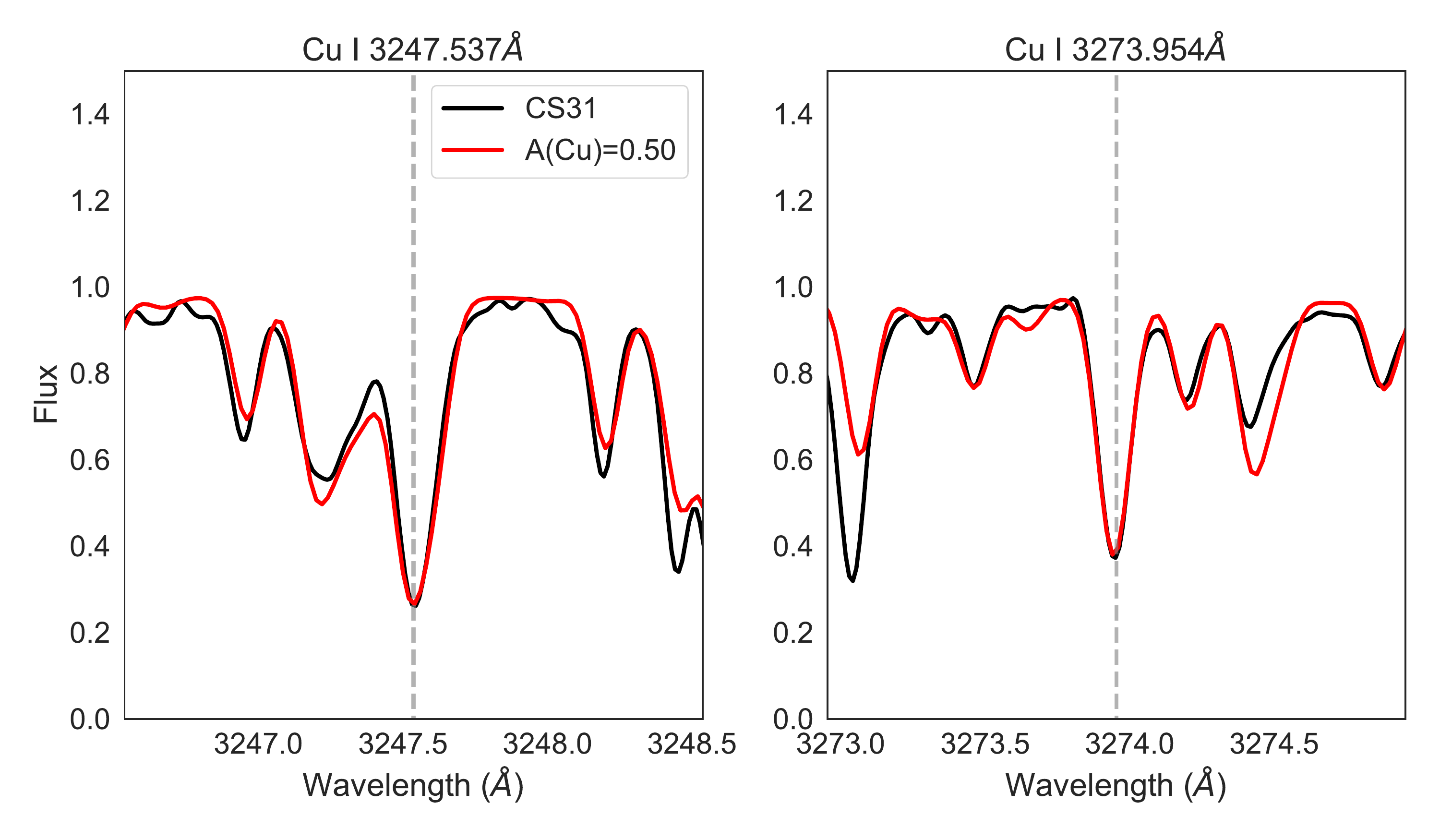}
    \caption{ Best fit for the Cu~I 3247.537 and 3273.954  {\rm \AA} lines, computed with [Cu/Fe]\,$=$\,$-$0.79 (including hyperfine structure).}
    \label{Cu}
\end{figure}

Our fits to the V and Cu lines are shown in Figs.~\ref{V} and \ref{Cu}.
From their large sample of metal-poor stars, Ou et al. (2020) found a discrepancy between abundance estimates from V~I and V~II, with mean values of [V~I/Fe]\,$=$\,$-$0.10 and [V~II/Fe]\,$=$\,$+$0.13. The difference is thought to arise from NLTE effects in the V~I lines, so for CS~31082-001 we used the V~II lines.

\begin{figure}
    \centering
    \includegraphics[width=3.3in]{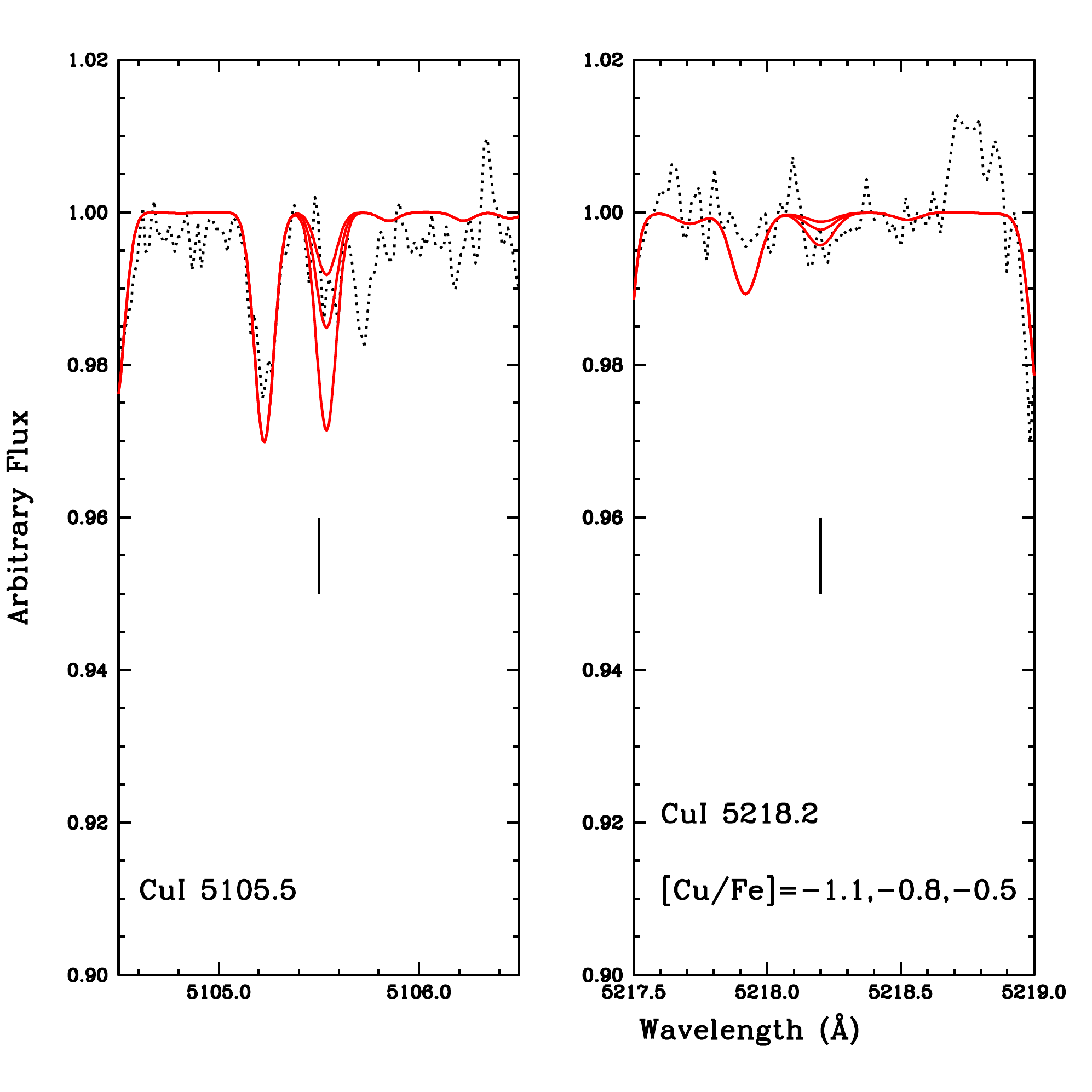}
    \caption{Best fit to the Cu~I 5105.5 and 5218.2 {\rm \AA} lines, computed with [Cu/Fe]\,$=$\,$-$0.80.}
    \label{newcu}
\end{figure}

For Cu, Bonifacio et al. (2010) found a discrepancy between the abundances for turn-off stars from the multiplet~1 line in the UV at 3247.5~\AA\ and multiplet~2 at 5105.5 {\rm \AA}. To see if the same discrepancy is present in CS~31082-001, we also fitted the Cu\,I 5105.5 and 5218.2 {\rm \AA} lines (see Fig.~\ref{newcu}) taking into account the HFS from Ernandes et al. (2018). In this case the UV and optical abundances agree.

\begin{table}
\caption{Comparison of our derived abundances  with results from Hill et al. (2002). Solar  abundances are from Grevesse \& Sauval (1998).}
\label{comparison} 
\centering                  
\begin{tabular}{lccccc} 
\hline\hline             
Species   &  Z  & A(X)$_{\odot}$ & [X/Fe]$_{\rm Hill}$ & [X/Fe]$_{\rm present}$ & A(X)  \\
\hline  
Be\,II & 4  & 1.15     & \ldots & $-$0.90 &  $-$2.4 \\
\hline
Sc\,II  & 21 & 3.17    & +0.02   & +0.01   & 0.28 \\
Ti\,I   & 22 & 5.02    & +0.25   & +0.28   & 2.40 \\
Ti\,II  & 22 & 5.02    & +0.31   & +0.28   & 2.40 \\
V\,II   & 23 & 4.00    & \ldots  & +0.15   & 1.25 \\ 
Cr\,I   & 24 & 5.67    & $-$0.34 & $-$0.34 & 2.43 \\
Mn\,II  & 25 & 5.39    & $-$0.35 & $-$0.35 & 2.14 \\
Co\,I   & 27 & 4.92    & +0.27   & +0.26   & 2.28 \\
Ni\,I   & 28 & 6.25    & +0.03   & +0.02   & 3.37 \\
Cu\,I   & 29 & 4.21    & \ldots  & $-$0.81 & 0.50 \\ 
Zn\,I   & 30 & 4.60    & +0.19   & +0.18   & 1.88 \\ 
Ge\,I   & 32 & 3.41    & $-$0.55 & \ldots  & \ldots \\

\hline
\hline                          
\end{tabular}
\end{table}

\section{Chemical evolution models}
To investigate our estimated V and Cu abundances in the context of nucleosynthesis, we have employed the disk/halo chemical-evolution model from Lanfranchi \& Friaça (2003, hereafter LF03), which has been used to investigate the origin of damped Lyman-$\alpha$ systems in galaxy disks and of Lyman-limit systems in galactic halos (Viegas et al. 1999). It is assumed that the infall of gas from the halo into the Milky Way disk feeds the disk star formation and that the star-formation rate (SFR) follows power laws on mass surface density and gas fraction. The  formalism and basic equations of the model can be found in LF03. An exponential mass surface density profile is adopted for the disk, $\sigma \propto e^{-r/r_{G}}$, with $r_G$ = 2.6 kpc (Boissier \& Prantzos 2000). In the adopted SFR law, the specific SFR (i.e. the inverse of the local star-formation timescale) depends on the gas density  as $\nu \propto \rho^{1/2}$. The normalizations of the SFR are $\tilde\nu_H=0.2$ Gyr$^{-1}$ for the initial halo and $\tilde\nu_D=0.5$ Gyr$^{-1}$ for the disk at the age of the Galaxy ($t_G=13.5$ Gyr) and at the solar Galactocentric distance ($r_{\odot}=8$ kpc). 

The history of gas infall into the Galactic disk presents two phases, in line with scenarios such as the two-infall model (Chiappini et al. 2001). The infall rate into the disk is assumed to decline with time as $e^{-t/\tau _D}$. The first phase of the infall is rapid, with the infall timescale $\tau _D=$ 1 Gyr, from $t=0$ until $t=1$ Gyr. In the late infall, $\tau _D$ is given by $\tau _D$ (r) = 1 + 7/6(r$-$2),  for 2 $\leq $  r  $\leq $ 8 kpc, and  $\tau _D$ (r) = 8 + 0.5(r$-$8),  for r $>$ 8 kpc . The inner and the outer boundaries of the disk are set  at 2 and 18 kpc. At these radii, $\tau _D=$ 1 and 13 Gyr, respectively, and $\tau _D=$ 8 at $r_{\odot}$. In contrast to Chiappini et al. (2001), we do not consider a threshold for star formation, fixed by adjusting its value in order to reproduce the observations, but we included inhibition of star formation by considering fundamental processes (see LF03 and Friaça \& Terlevich 1998) 

With respect to the nucleosynthesis prescriptions of our model, we adopt metallicity-dependent yields from core-collapse supernovae (SNe II), high explosion-energy hypernovae, type Ia supernovae (SNe Ia), and intermediate-mass stars (IMS) (for more details, see  Friaça \& Barbuy 2017). The SN II yields are adopted from WW95. For low metallicities (Z<0.01Z$_{\odot}$), we also considered the yields from  hypernovae (Umeda \& Nomoto 2002, 2003, 2005; Nomoto et al. 2006, 2013). The yields of SNIa resulting from Chandrasekhar mass white dwarfs are taken from Iwamoto et al. (1999), specifically models W7 (progenitor star of initial metallicity Z\,$=$\,Z$_{\odot}$) and W70 (zero initial metallicity). The yields for IMS (0.8-8 M$_{\odot}$) with initial Z\,$=$\,0.001, 0.004, 0.008, 0.02, and 0.4 are from van den Hoek \& Groenewegen (1997) (variable $\eta_{AGB}$ case).

\subsection{Abundances and chemical-evolution models for V}

The upper panel of Fig.~\ref{VCuplot} shows V abundance estimates derived by Ou et al. (2020) for the sample of halo stars from Roederer et al. (2014), and the results from Ishigaki et al. (2013) for halo and thick disk stars; our best fit abundance, derived from the V~II 3951.96\,\AA\ line in CS~31082-001 is shown by the red circle.

Vanadium is synthesized in explosive Si-burning during core-collapse supernovae (SNe). Its abundances are known to be underpredicted by chemical evolution models (at all metallicities) when compared to observations (WW95, Kobayashi et al. 2006, 2020), as highlighted by the model from Kobayashi et al. (2020) in the upper panel of Fig.~\ref{VCuplot} compared to the observational results.

Multidimensional effects could increase the abundances of Sc, Ti and V (Maeda \& Nomoto 2003; Tominaga 2009). To mimic these effects, the K15 model of Kobayashi et al. (2020) considered a 0.3 dex enhancement in the V yields. Indeed, Timmes et al. (1995) pointed out that the V yields of WW95 should be increased by a factor of $\sim$3 to reproduce the observational data. Therefore, in our chemical-evolution models, we applied this factor to the V yields of WW95 from $Z$\,$=$\,$Z{_\odot}$ to 0.01\,$Z{_\odot}$. Including an enhancement by a constant factor accounts for the flat behaviour of [V/Fe] for [Fe/H]\,$>$\,$-$2. However, as V varies in lockstep with iron, the models do not reproduce the high [V/Fe] ratios seen in extremely metal-poor and very metal-poor stars, including CS~31082-001.

Core-collapse SNe release large amounts of energy as neutrinos ($>$\,10$^{53}$ erg). The interaction of neutrinos with matter represents an additional nucleosynthetic source (WW95, Heger et al. 2005, Yoshida et al. 2008). In very low metallicity stars, neutrino processes contribute significantly to the production of odd-Z elements such as V, Mn, Sc, K, F and B (Yoshida et al. 2008, Kobayashi et al. 2011a). Following the prescriptions of Yoshida et al. (2008), our models therefore also include enhancement of V yields by neutrino processes during the SN explosion. These phenomena can be important at very low metallicities but would be ineffective for the yields of hypernovae. The total neutrino energy E$_\nu$ released when the core of a massive star collapses to form a neutron star is a free parameter. In our models we adopted the standard case of E$_\nu$\,$=$\,3\,$\times$\,10$^{53}$ erg from Yoshida et al. (2008), which corresponds to the gravitational binding energy of a 1.4M$_\odot\ $ neutron star (Lattimer \& Prakash 2001). We also consider the cases of a larger neutrino energy E$_\nu$\,$=$\,9\,$\times$\,10$^{53}$ erg, and the standard case of E$\nu$ = 3 $\times$ 10$^{53}$ erg but with a neutrino temperature of T$_\nu$\,$=$\,8 Mev (as assumed by WW95). As shown by Fig.~\ref{VCuplot}, inclusion of the neutrino processes better reproduces the rising trend of [V/Fe] for [Fe/H]\,$<$\,$-$2.0 in the results from Ou et al. (2020) as well as our result for CS~31082-001.

We have considered neutrino processes as an additional nucleosynthetic source of vanadium. Although this effect
reproduces the rise of [V/Fe] for [Fe/H] $\lesssim -2.5$ , it falls
short of the highest observed values of [V/Fe] around +1.5, obtained
by Ou eu al. (2020). In this context it is worth considering the possibility of
observational errors. Following Ou et al.
(2020) the vanadium abundance for CS~31082-001 was derived from the V~II lines. The highest value of [V/Fe] derived by Ou et
al. (2020) using the V~II line is 1.63 for G238$-$030. If using the
V I line instead, the derived [V/Fe] is 1.46. On the theoretical side, it could
be that other processes are at work, e.g., the above mentioned
multidimensional effects. Even considering the sole neutrino process,
our models have been conservative, and higher neutrino energies would
increase the [V/Fe] ratio.

\subsection{Abundances and chemical evolution models for Cu}

Copper is produced both in the alpha-rich freeze-out as a primary element (Sukhbold et al. 2016) and in the weak-s process in massive stars as a secondary element. The iron-peak elements are mainly formed during explosive oxygen and silicon burning in massive supernovae (WW95). For larger values of the neutron fraction, the main products of silicon burning are completed. On the other hand, if the density is low and the supernova envelope expansion is fast, $\alpha$ particles will be frozen and not captured by the heavier elements (Woosley et al. 2002). The trend of Cu abundance with the metallicity [Fe/H] could reveal the relative efficiencies of these two contributions.

The secondary-like behaviour of Cu has led Sneden et al. (1991) to suggest that it could mainly be attributed to the weak s-process. However, as more data has been accumulated and chemical evolution models have been tested, a more complex picture has emerged (Mishenina et al. 2002, Kobayashi et al. 2006, 2011b, 2020). The relation of [Cu/Fe] versus [Fe/H] for $-$2 < [Fe/H] < $-$0.5 is well described by a secondary process, but if we consider a wider range in metallicity, from [Fe/H]  $\sim$ $-$4 to $\sim$ 0, the behavior of [Cu/Fe] is more complex than that expected from purely primary or secondary processes. The general curve has a wavy shape, and for  [Fe/H]  < $-$2.5, [Cu/Fe] increases or, at least, reaches a plateau, indicating the presence of a primary process. In fact, the Cu content of extremely metal-poor stars is basically determined by explosive nucleosynthesis in massive stars, and by hypernovae in particular. On the other hand, the form of the trend of [Cu/Fe] with [Fe/H] for higher metallicities, tells us about the role of longer lived sources of Cu enrichment. At [Fe/H] $\sim$ $-$1, [Cu/Fe] shows another plateau, which is explained by SNe Ia.  In addition to that, besides an origin in massive stars, Cu is also formed through the s-process acting in AGB stars. However, this contribution is negligible at [Fe/H]  $\sim$ $-$2 and amounts to only 0.03 dex at [Fe/H] = 0 (Kobayashi et al. 2020).

\begin{figure*}
    \centering
    \includegraphics[ trim=0cm 5cm 0cm 3cm,width=5.0in]{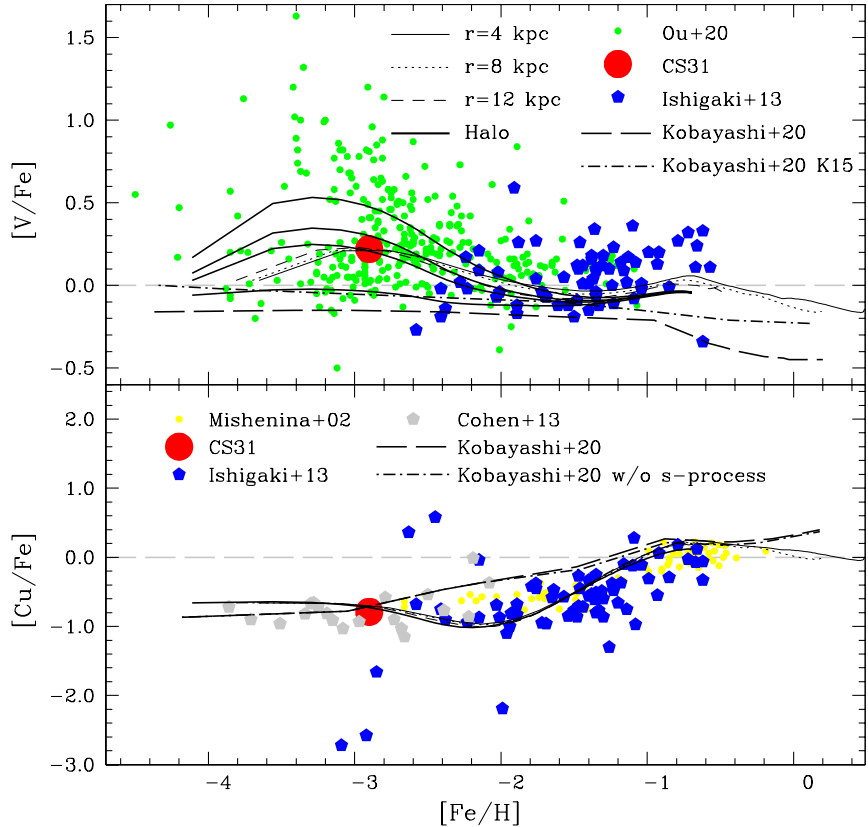}
    \caption{{\em Upper panel:} Relation of [V~II/Fe] vs. [Fe/H]. Symbols: red dots:
present work; green dots: Ou et al. (2020), blue pentagons: Ishigaki
et al. (2013). Our chemical evolution model for the halo/disk (LF03)
is shown for several Galactocentric radii and for the halo (thick
solid lines). Our fiducial model includes neutrino processes with
E$_{\nu}=3\times 10^{53}$ erg. The various curves for the halo
correspond to different assumptions for the neutrino process. From
bottom to top: 1) no neutrino process; 2) the fiducial model
(E$_{\nu}=3\times 10^{53}$ erg); 3)  E$_{\nu}=3\times 10^{53}$ erg and
 T$_{\nu}$ = 8 Mev;  4) E$_{\nu}=9\times 10^{53}$ erg. As we can see
from the comparison of the LF03 model with the data, inclusion of the neutrino process
could account for the high [V/Fe] ratios of extremely metal-poor and
very metal-poor stars. The solid long-dashed line refers to the
predictions of the fiducial  model of Kobayashi et al. (2020), while the
K15 model from Kobayashi et al. (2020), which includes an 0.3
dex enhancement of the V yields, is also shown (dotted-dashed line). {\em Lower panel:}
Relation of [Cu~I/Fe] vs. [Fe/H]. Symbols: red dots: present work; yellow
dots: Mishenina et al. (2002); blue pentagons: Ishigaki et al. (2013);
gray pentagons: Cohen et al. (2013). The thick solid line refers to
the halo in the LF03 model and the thin lines to the disk in the model
at the radii r= 4, 8 and 12 kpc. The neutrino process follows the
fiducial case (E$_{\nu}=3\times 10^{53}$ erg). The LF03 model  not
only reproduces the CS 31082-001 data point but also predicts the wave
shape of the [Cu/Fe] versus [Fe/H] curve. The resulting evolution of
Cu abundance predicted by the model of Kobayashi et al. (2020) for
the solar neighborhood  is given by the black long-dashed line, and the
dotted-dashed line refers to the model without production of Cu by
the s-process in AGB stars.}
\label{VCuplot}
\end{figure*}

In the lower panel of Fig.~\ref{VCuplot} we show the Cu abundances derived by Mishenina et al. (2002), Cohen et al. (2013), and Ishigaki et al. (2013) for halo and thick disk stars and our best fit abundance to the Cu~I 3247.54 and 3273.95\,\AA\ lines of CS 31082-001 compared to the predictions of the LF03 model used in this work. The data points from Ishigaki et al. (2013) are systematically below those from Mishenina et al. (2002) and, although their scatter is large, they suggest a sinuous behavior of [Cu/Fe] with [Fe/H], as predicted by the LF03 model. The abundances we derived for CS~31082-001 (red circle) are well fitted by our model. The plateau of [Cu/Fe] at [Fe/H] < 2.5 appears to mainly be the result of hypernovae.

The thick solid line in the lower panel of Fig.~\ref{VCuplot} refers to the halo in our
model and the thin lines to the disk at the radii r\,$=$\,4, 8 and 12 kpc.
The neutrino process follows the fiducial case (E$_{\nu}=3\times
10^{53}$ erg), but it has little impact on the Cu abundances.
The results in Fig.~\ref{VCuplot} also include the predictions for Cu abundances from Kobayashi et al. (2020), whose models include AGBs, core-collapse SNe and type Ia SNe. This model accounts for the [Cu/Fe] ratio of CS~31082-001 but overestimates the Cu abundances around [Fe/H] = $-$2.0. To assess the effect of the s-process in AGB stars on Cu production, their model without the AGB contribution is also plotted (dotted-dashed line in Fig.~\ref{VCuplot}). This process contributes with only a very modest increase of Cu near [Fe/H] = 0.0, and removing the AGB contribution does not alleviate the overestimate of [Cu/Fe] at  [Fe/H] $\sim$ $-$2.0.

\subsection{Abundances: Summary}
\begin{figure*}
    \centering
    \includegraphics[width=6.0in]{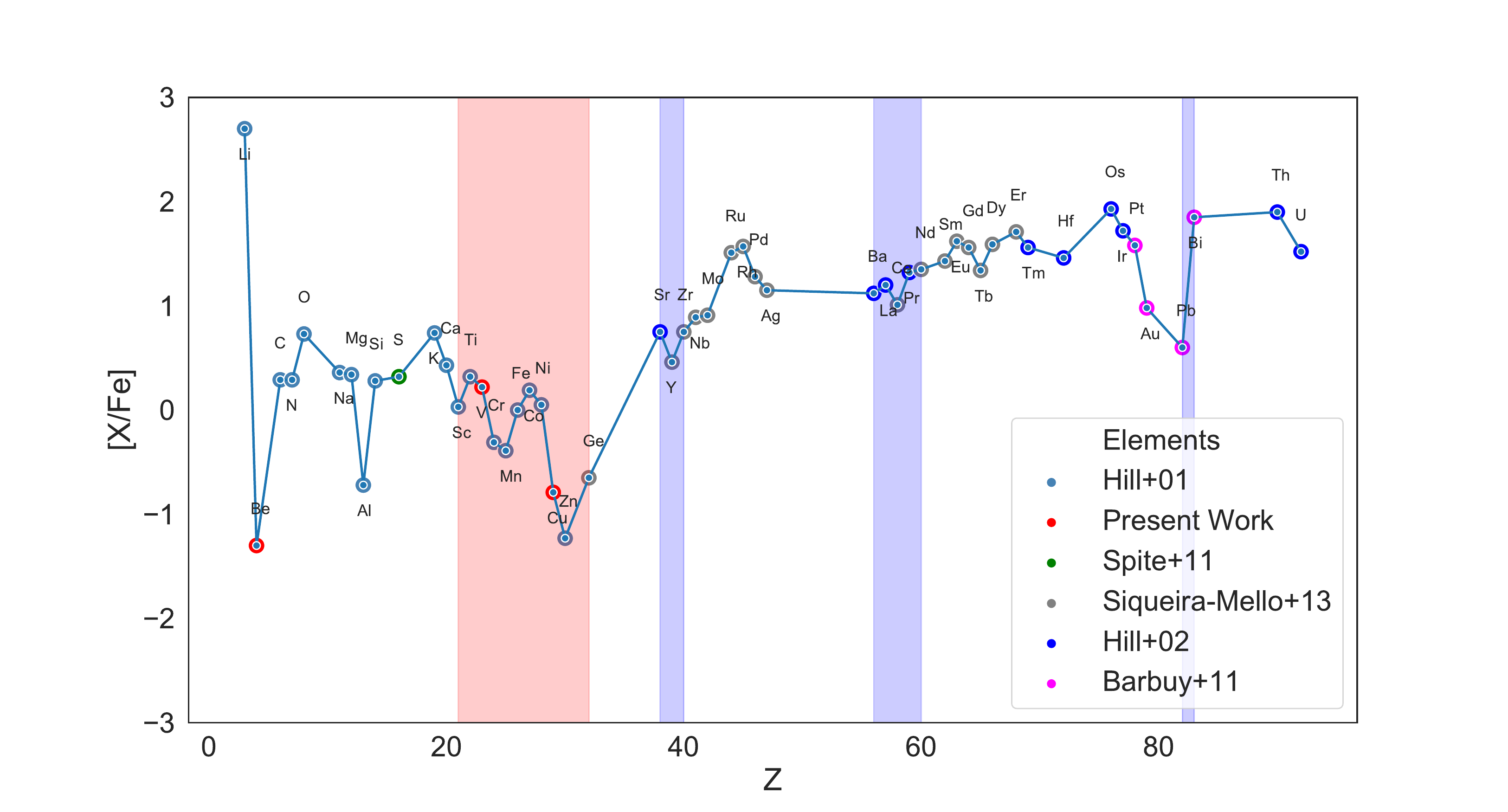}
    \caption{Elemental abundances of CS~31082-001, with the symbol colours showing the source of the result, as indicated in the legend. The red shaded region highlights the iron-peak elements, with the blue regions highlighting the first, second and third peaks of  s-process elements}. 
    \label{zplot}
\end{figure*}

The iron-peak abundance pattern of CS~31082-001 is shown in Fig.~\ref{zplot}, including results from Hill et al. (2001, 2002), Cayrel et al. (2004), Plez et al. (2004),  Spite et al. (2005, 2011), Barbuy et al. (2011), Siqueira-Mello et al. (2013), and our new results for  V and Cu.

\begin{figure}
    \centering
    \includegraphics[width=3.3in]{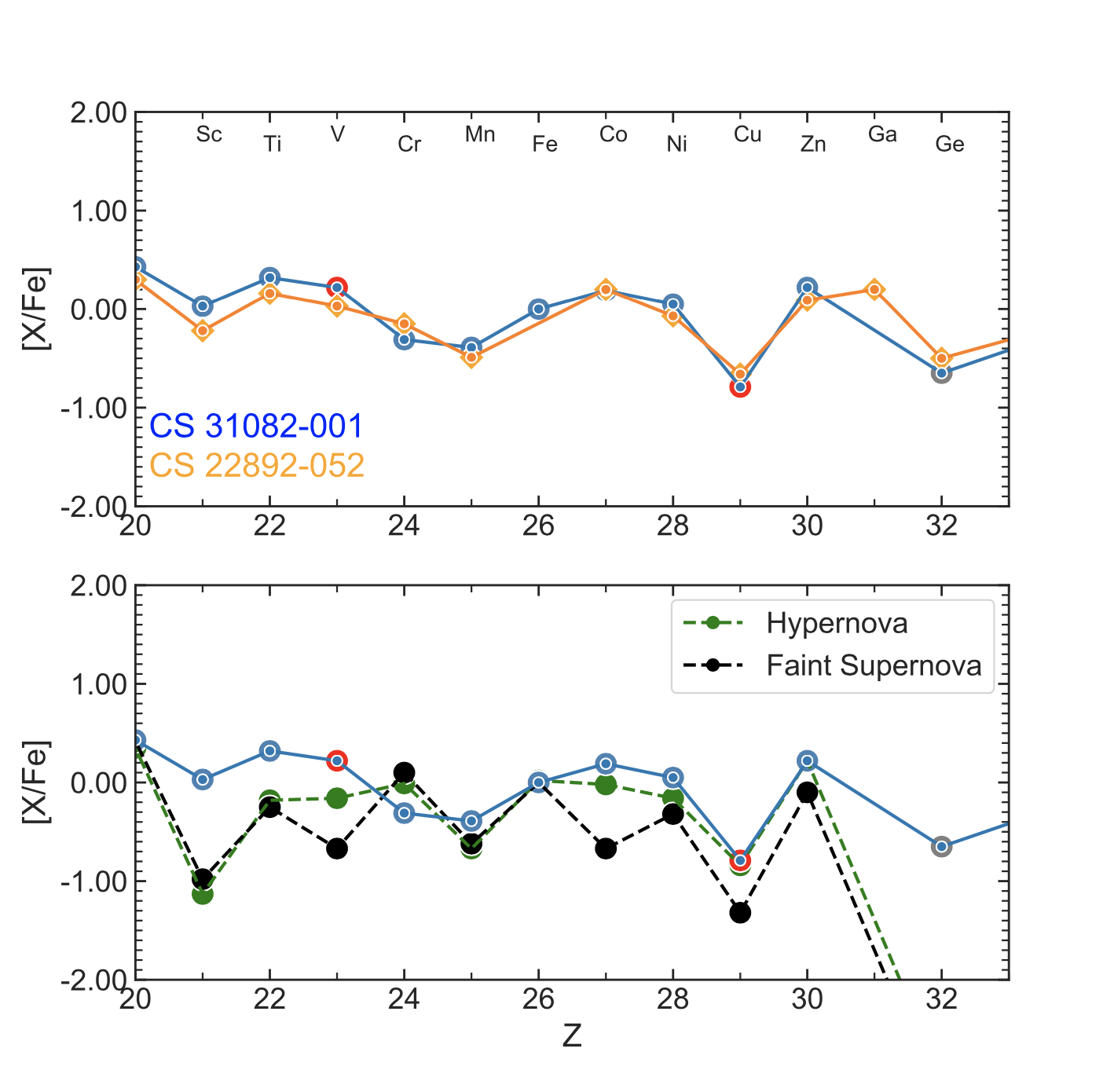}
    \caption{{\em Upper panel:} Elemental abundances of CS~31082-001. Symbols: red circles are from the present work; blue circles, Hill et al. (2002); gray circles, Siqueira-Mello et al. (2013). The orange diamonds are the elemental abundances of CS~22892-052 from Sneden et al. (2003). {\em Lower panel:} Iron-peak abundances of CS~31082-001 with the same symbol colours as the upper panel. The predicted abundance yields from for Nomoto et al. (2013) for faint supernovae and hypernovae are shown by the green and black dashed lines, respectively.} 
    \label{ironplot}
\end{figure}

In the upper panel of Fig.~\ref{ironplot} we compare the iron-peak abundances in CS~31082-001 with those for the similarly r-process-rich star CS~22892-052 (Sneden et al. 2003). These two halo metal-poor stars appear to have not only the same abundance patterns for the neutron-capture elements but also have the same pattern for the iron-peak elements, indicating that the same kind of process produced the iron-peak elements in both stars. To confirm the
chemical evolution models developed above, in the lower panel of Fig.~\ref{ironplot} we show the abundances of iron-peak elements compared with the predicted SNe yields from Nomoto et al. (2013) for a faint supernova (black, dashed line) and a hypernova (green, dashed line). The latter is a better fit to the iron-peak elements (e.g. for Co, Ni, Cu, and Zn) but both models underestimate the production for Sc, Ti, V, Mn, and Ge, and overestimate the yield for Cr. This is the reason why 
we combined yields from hypernovae and added neutrino processes in our chemical evolution models. In particular for V, it adjusts the lower [V/Fe]\,$=$\,$-$0.1 yield to reach [V/Fe]\,$=$\,+0.3.

The underestimates in the production of the Sc, V, Mn, and Co in the hypernovae models can be explained by the $\nu-$process not being included in the model from Nomoto et al. (2013). According to Yoshida et al. (2008) the hypernovae models can be improved by adding the $\nu-$process -- especially for the odd-Z elements where the neutrino process plays a significant role. 
One of these elements is V, which is mainly synthesized in the incomplete Si-burning region as $^{51}$Mn. However, in the hypernovae models it is synthesized in the complete Si-burning region along with the $\nu-$process from $^{52}$Fe in the incomplete Si-burning region.
The [Ge/Fe] abundances are significantly underestimated by the faint SNe and the hypernovae models. However, it is also underproduced in the models from Wanajo (2007), as described by Siqueira-Mello (2013) in a previous analysis of CS~31082-001, where they concluded that the r-process considered by Wanajo (2007) cannot explain the observed Ge abundance. On the other hand, the model of high entropy winds by Farouqi et al. (2010), with an electron abundance of Y$_{e}$ = 0.498, can explain the abundance of the trans-iron elements Ge, As, Se, Mo, and Ru (Peterson et al. 2020).

\section{Summary}

We have used the near-UV region of high-resolution spectra from VLT-UVES and {\em HST}-STIS to estimate Be, V and Cu abundances in the metal-poor star CS~31082-001 for the first time. Beryllium in CS~31082-001 appears to be low, with A(Be)\,$=$\,$-$2.4 to $-$2.5, and not following the linear trend seen for unmixed, metal-poor stars. This is expected for a cool red giant, due to diffusion effects and consequent destruction of the light elements (Li and Be) in the atmospheric layers.

Our estimated abundances of the iron-peak elements V and Cu agree well with other data in the literature, and we present new chemo-dynamical models for these elements. We can only reproduce the
observed V abundance by adding an extra nucleosynthesis source to the models to include the interaction of neutrinos with matter. For Cu we found that the behaviour of [Cu/Fe] for metal-poor stars
confirms a decreasing Cu-to-Fe abundance with decreasing metallicity. As pointed out by Sukhbold et al. (2016), Co and Cu are both produced in the alpha-rich freezeout as primary elements, and as well in the weak-s process in massive stars, in the latter case with a secondary element behaviour (see also Barbuy et al. 2018). A dominant weak-s process origin is therefore confirmed for copper.

We have also investigated the abundances estimated for iron-peak elements from near-UV diagnostics compared with published values from longer-wavelength lines in the visible, finding good agreement. This validation is important in the context of future near-UV observations of metal-poor stars with the CUBES instrument. In a future paper we will investigate the detectability of heavy elements in the UV region in CS~31082-001, to further investigate the future opportunities with CUBES in studies of stellar nucleosynthesis.


\section*{Acknowledgements}
HE acknowledges a CAPES PhD fellowship, and a CAPES-PRINT fellowship in support of his one year of study at the University of Edinburgh. BB acknowledges partial financial support from FAPESP, CNPq and CAPES-Financial code 001. HE and CJE also acknowledge support from the Global Challenges Research Fund (GCRF) from UK Research and Innovation. Based on observations gathered at the Very Large Telescope at Paranal Observatory, Chile, under Program ID 165-N-0276.
\section*{Data Availability}

The inclusion of a Data Availability Statement is a requirement for articles published in MNRAS. Data Availability Statements provide a standardised format for readers to understand the availability of data underlying the research results described in the article. The statement may refer to original data generated in the course of the study or to third-party data analysed in the article. The statement should describe and provide means of access, where possible, by linking to the data or providing the required accession numbers for the relevant databases or DOIs.







\appendix

\section{Atomic data }

In Table \ref{lines1} we summarise the 
 magnetic dipole A-factor and electronic quadrupole B-factor
 hyperfine constants used to compute the hyperfine structure (HFS) of odd-Z
 elements. The list of lines from the inclusion of HFS for Co~I 3412.34, 3412.63,
 and 3449.16\,\AA\ are given in Table~\ref{hfsCo3}, for 
 Co~I 3529.03, 3842.05, and 3845.47\,\AA\ in Table~\ref{hfsCo2}, and for
 Cu~I 3247.53 and 3273.95\,\AA\ in Table~\ref{hfsCu2}.
\begin{table*}
\caption{Hyperfine constants A and B for Sc~II (Villemoes et al. 1992; Kurucz, 1993); V~II (Wood et al. 2014); Mn~II (Den Hartog et al. 2011); Co~I (Pickering 1996); Cu~I (Biehl, 1976; Kurucz, 1993). For transitions where the electric quadrupole constants were not available we adopted the B-factor as zero.}             
\label{lines1}      
\centering          
\begin{tabular}{lc@{~~~}c@{~~~}c@{~~~}c@{~~~}c@{~~~}c@{~~~}c@{~~~}c@{~~~}c@{~~~}c@{~~~}c@{~~~}c@{~~~}c} 
\noalign{\smallskip}
\hline\hline    
Species & $\lambda$ & \multicolumn{6}{c}{Lower level} & \multicolumn{6}{c}{Upper level} \\

& & \pp Config.\pp & \pp J\pp& A & A & B & B & Config. & J & A & A & B & B \\
 & ({\rm \AA}) & &  & (mK)& (MHz) & (mK)& (MHz) & &  &  (mK)&  (MHz) & (mK)& (MHz)  \\
 
\noalign{\vskip 0.1cm}
\noalign{\hrule\vskip 0.1cm}
\noalign{\vskip 0.1cm}
$^{45}$Sc~II & 3576.34  & 3d4s 3D & 2.0 & 507.67 & 15219.5665 & $-$34.7 & $-$1040.28  & 3d4p 3D & 2.0 & 125.3 & 758.475 & 10.0 & 299.7925 \\
$^{45}$Sc~II & 3590.47 & 3d4s 3D & 3.0 &654.8  & 19630.4141 & $-$63.0 & $-$1888.6929 & 3d4p 3D & 2.0 & 125.3 & 758.475 & 10.0  & 299.7592 \\
\noalign{\vskip 0.1cm}
\noalign{\hrule\vskip 0.1cm}
\noalign{\vskip 0.1cm}
$^{51}$V~II & 3517.299 & (4F)4s a3F & 4.0 & -2.9  & -86.9398 & \ldots & \ldots &  (4F)4p z5D & 3.0 & \ldots & \ldots & \ldots & \ldots\\

$^{51}$V~II & 3545.196 & (4F)4s a3F & 3.0 & 6.0  & 179.8755 & \ldots & \ldots &  (4F)4p z3D & 2.0 & \ldots & \ldots & \ldots & \ldots\\

$^{51}$V~II & 3715.464 & d4 a3H  & 6.0 & ---  & --- & \ldots & \ldots &  (4F)4p z3G & 5.0 & \ldots & \ldots & \ldots & \ldots\\

$^{51}$V~II & 3951.96 & d4 a3P & 2.0 & 0.0  & 0.0 & \ldots & \ldots &  (4F)4p z3D & 3.0 & \ldots & \ldots & \ldots & \ldots\\
\noalign{\vskip 0.1cm}
\noalign{\hrule\vskip 0.1cm}
\noalign{\vskip 0.1cm}  

$^{55}$Mn~II & 3441.99& d6 a5D  & 4.0 &74.0 & 2218.4646 & \ldots & \ldots &  (6S)4p z5P & 3.0 & $-$150.3 & $-$4505.8818 & 85.0 & 2548.2363 \\
$^{55}$Mn~II & 3460.32& d6 a5D  & 3.0 &5.8 & 173.8797 & $-$71.0 & $-$2129.5269&  (6S)4p z5P & 2.0 & $-$310.7 &$-$9314.5527 & $-$87.0 & $-$2608.1948 \\
$^{55}$Mn~II & 3482.90& d6 a5D  & 2.0 &$-$35.0 & $-$1049.2738 & $-$40.0 & $-$1199.1700&  (6S)4p z5P & 2.0 &$-$310.7& $-$9314.5527 & $-$87.0 & $-$2608.1948 \\
$^{55}$Mn~II & 3488.68& d6 a5D  & 1.0 &$-$59.0 & $-$1768.7758 & $-$53.0& $-$1588.9004&  (6S)4p z5P & 1.0 & $-$737.0 & 22094.7090 & 9.0 & 269.8133 \\
$^{55}$Mn~II & 3495.83& d6 a5D  & 0.0 &0.0 & 0.0  & \ldots & \ldots &  (6S)4p z5P & 1.0 & $-$737.0& 22094.7090 & 9.0 & 269.8133 \\
$^{55}$Mn~II & 3497.53& d6 a5D  & 1.0 &$-$59.0 & $-$1768.7758  & $-$53.0 &$-$1588.9004&  (6S)4p z5P & 2.0 & $-$310.7 & $-$9314.5527 & $-$87.0 & $-$2608.1948 \\

\noalign{\vskip 0.1cm}
\noalign{\hrule\vskip 0.1cm}
\noalign{\vskip 0.1cm}
$^{59}$Co~I & 3412.34 & (3F)4s b4F  & 3.5 & 22.3127  & 668.9181 & $-$2.643 & $-$79.2352 & (3F)4p y2G & 4.5 & 14.73 & 441.5943 & 0.0 & 0.0 \\
$^{59}$Co~I & 3412.63 & d7s2        & 4.5 & 15.01984 & 450.2835 & 4.644  & 139.2236 & 4F)4sp z4D & 3.5 & 25.05 & 750.9802 & 3.0 & 89.9378 \\
$^{59}$Co~I & 3449.16 & (3F)4s b4F  & 2.5 & 18.7524  & 562.1829 & $-$1.828 & $-$54.8021 & (3F)4p y4G & 2.5 & 26.85 & 804.9429 & $-$4.0 & $-$119.9170 \\
$^{59}$Co~I & 3529.03 & d7s2 a4F    & 2.5 & 20.4591  & 613.3486 & 2.253  & 67.5433 & 4F)4sp z4G & 3.5 & 14.95 & 448.1898 & 5.0 & 149.8963 \\
$^{59}$Co~I & 3842.05 & (3F)4s a2F  & 3.5 & 13.01    & 390.0301 & $-$5     & $-$149.8963 & 4F)4sp z2D & 2.5 & 15.4  & 461.6805 & \ldots & \ldots \\
$^{59}$Co~I & 3845.47 & (3F)4s a2F  & 3.5 & 13.01    & 390.0301 & $-$5     & $-$149.8963 & (3F)4p y2G & 4.5 & 14.73 & 441.5943 & 0.0 & 0.0 \\

\noalign{\vskip 0.1cm}
\noalign{\hrule\vskip 0.1cm}
\noalign{\vskip 0.1cm}
$^{63}$Cu~I & 3247.53 & 4s 2S  & 0.5 &194 & 5815.9746 & 0.0 & 0.0 & 4p 2P & 1.5 & 6.5 & 194.685  & $-$0.96 & $-$28.78 \\
$^{63}$Cu~I & 3273.95 & 4s 2S  & 0.5 &194 & 5815.9746  & 0.0 & 0.0 & 4p 2P & 0.5 & 6.5 & 194.685  & $-$0.96 & $-$28.78 \\
$^{65}$Cu~I & 3247.53 & 4s 2S  & 0.5 &208 & 6235.6841 & 0.0 & 0.0 &  4p 2P & 1.5 & 6.96 & 208.66 & $-$0.86 & $-$25.78 \\
$^{65}$Cu~I & 3273.95 & 4s 2S  & 0.5 &208 & 6235.6841 & 0.0 & 0.0 & 4p 2P & 0.5 & 6.96 & 208.66 & $-$0.86 & $-$25.78 \\
\hline                  
\end{tabular}
\end{table*}

\begin{table*}
\caption{Calculated hyperfine structure for the Co~I 3412.34, 3412.63, and 3449.16\,\AA\ lines.}
\label{hfsCo3}
\centering
\begin{tabular}{ccccccccccccc}
\hline
\noalign{\smallskip}
\multicolumn{3}{c}{3412.34\AA;  $\chi$\,$=$\,0.5136 eV} && \multicolumn{3}{c}{3412.63\AA; $\chi$\,$=$\,0.00 eV}   && \multicolumn{3}{c}{3449.16\AA; $\chi$\,$=$\,0.5815  eV}  & \\
\multicolumn{3}{c}{log gf(total) = 0.030} && \multicolumn{3}{c}{log gf(total) = $-$0.780} && \multicolumn{3}{c}{log gf(total) = $-$0.090}   & \\
\noalign{\smallskip}
\noalign{\smallskip}
\cline{1-3} \cline{5-7} \cline{9-11} \\
$\lambda$ (\AA) & log gf & iso && $\lambda$ (\AA) & log gf & iso && $\lambda$ (\AA) & log gf &iso \\
\noalign{\smallskip}
\cline{1-3} \cline{5-7} \cline{9-11} \\
3412.332 &   $-$1.776 & 59 && 3412.643 & $-$2.5861 & 59 &&  3449.171 & $-$1.741 & 59 & \\
3412.334 &   $-$1.709 & 59 && 3412.640 & $-$2.519  & 59 &&  3449.164 & $-$1.486 & 59 & \\
3412.331 &   $-$1.513 & 59 && 3412.634 & $-$3.218  & 59 &&  3449.176 & $-$1.486 & 59 & \\
3412.340 &   $-$2.408 & 59 && 3412.643 & $-$2.323  & 59 &&  3449.169 & $-$3.395 & 59 & \\
3412.336 &   $-$1.513 & 59 && 3412.637 & $-$2.323  & 59 &&  3449.159 & $-$1.287 & 59 & \\
3412.331 &   $-$1.309 & 59 && 3412.629 & $-$3.063  &    &&  3449.176 & $-$1.287 & 59 & \\
3412.344 &   $-$2.25  & 59 && 3412.642 & $-$2.119  &    &&  3449.166 & $-$2.152 & 59 & \\
3412.339 &   $-$1.40  & 59 && 3412.634 & $-$2.218  &    &&  3449.153 & $-$1.223 & 59 & \\
3412.332 &   $-$1.13  & 59 && 3412.622 & $-$3.063  &    &&  3449.175 & $-$1.223 & 59 & \\
3412.350 &   $-$2.25  & 59 && 3412.640 & $-$1.949  &    &&  3449.162 & $-$1.434 & 59 & \\
3412.343 &   $-$1.36  & 59 && 3412.629 & $-$2.171  &    &&  3449.146 & $-$1.254 & 59 & \\
3412.334 &   $-$0.99  & 59 && 3412.614 & $-$3.160  &    &&  3449.173 & $-$1.254 & 59 & \\
3412.356 &   $-$2.35  & 59 && 3412.637 & $-$1.803  &    &&  3449.157 & $-$1.028 & 59 & \\
3412.347 &   $-$1.36  & 59 && 3412.623 & $-$2.177  &    &&  3449.139 & $-$1.435 & 59 & \\
3412.363 &   $-$2.55  & 59 && 3412.633 & $-$1.674  &    &&  3449.152 & $-$0.737 & 59 & \\
3412.353 &   $-$1.44  & 59 && 3412.615 & $-$2.250  &    &&   & & & \\
3412.341 &   $-$0.74  & 59    && 3412.594 & $-$3.762   &    && &&& \\
3412.370 &   $-$2.95  & 59    && 3412.628 & $-$1.558  &    && &&& \\
3412.358 &   $-$1.65  & 59    && 3412.607 & $-$2.461   &    && &&& \\
3412.345 &   $-$0.64  & 59    && 3412.622 & $-$1.453   &    && &&& \\
\noalign{\smallskip}
\hline
\end{tabular}
\end{table*}

\begin{table*}
\caption{Calculated hyperfine structure for the Co~I 3529.03, 3842.05, and 3845.47\,\AA\ lines.}
\label{hfsCo2}
\centering
\begin{tabular}{ccccccccccccc}
\hline
\noalign{\smallskip}
\multicolumn{3}{c}{3529.03\AA;  $\chi$\,$=$\,0.1744 eV} && \multicolumn{3}{c}{3842.05\AA; $\chi$\,$=$\,0.9227 eV}   && \multicolumn{3}{c}{3845.47\AA; $\chi$\,$=$\,0.9227  eV}  & \\
\multicolumn{3}{c}{log gf(total) = $-$0.880} && \multicolumn{3}{c}{log gf(total) = $-$0.770} && \multicolumn{3}{c}{log gf(total) = 0.010}   & \\
\noalign{\smallskip}
\noalign{\smallskip}
\cline{1-3} \cline{5-7} \cline{9-11} \\
$\lambda$ (\AA) & log gf & iso && $\lambda$ (\AA) & log gf & iso && $\lambda$ (\AA) & log gf &iso \\
\noalign{\smallskip}
\cline{1-3} \cline{5-7} \cline{9-11} \\
  3529.030 &  $-$2.686 & 59 && 33842.044& $-$2.566 & 59 &&  3845.480 & $-$1.796 & 59 & \\
  3529.028 &  $-$2.401 & 59 && 33842.046& $-$2.281 & 59 &&  3845.482 & $-$1.729 & 59 & \\
  3529.025 &  $-$2.656 & 59 && 33842.042& $-$2.536 & 59 &&  3845.478 & $-$1.533 & 59 & \\
  3529.033 &  $-$2.656 & 59 && 33842.051& $-$2.536 & 59 &&  3845.487 & $-$2.428 & 59 & \\
  3529.030 &  $-$2.213 & 59 && 33842.046& $-$2.093 & 59 &&  3845.482 & $-$1.533 & 59 & \\
  3529.025 &  $-$2.268 & 59 && 33842.039& $-$2.587 &    &&  3845.476 & $-$1.329 & 59 & \\
  3529.037 &  $-$2.707 & 59 && 33842.052& $-$2.148 &    &&  3845.489 & $-$2.273 & 59 & \\
  3529.032 &  $-$2.122 & 59 && 33842.045& $-$2.002 &    &&  3845.482 & $-$1.428 & 59 & \\
  3529.025 &  $-$2.013 & 59 && 33842.036& $-$2.712 &    &&  3845.473 & $-$1.159 & 59 & \\
  3529.042 &  $-$2.832 & 59 && 33842.054& $-$1.893 &    &&  3845.490 & $-$2.273 & 59 & \\
  3529.035 &  $-$2.100 & 59 && 33842.045& $-$1.980 &    &&  3845.482 & $-$1.381 & 59 & \\
  3529.026 &  $-$1.815 & 59 && 33842.033& $-$2.934 &    &&  3845.471 & $-$1.013 & 59 & \\
  3529.048 &  $-$3.054 & 59 && 33842.054& $-$1.695 &    &&  3845.491 & $-$2.370 & 59 & \\
  3529.039 &  $-$2.154 & 59 && 33842.043& $-$2.034 &    &&  3845.481 & $-$1.387 & 59 & \\
  3529.028 &  $-$1.651 & 59 && 33842.030& $-$3.344 &    &&  3845.468 & $-$0.884 & 59 & \\
  3529.054 &  $-$3.464 & 59 && 33842.054& $-$1.531 &    &&  3845.492 & $-$2.574 & 59 & \\
  3529.043 &  $-$2.350 & 59 && 33842.041& $-$2.230 &    &&  3845.479 & $-$1.460 & & \\
  3529.029 &  $-$1.510 & 59 && 33842.053& $-$1.390 &    &&  3845.464 & $-$0.768 & & \\
    &&                    &&          &        &    &&  3845.491 & $-$2.972 & &  \\
    &&                    &&          &        &    &&  3845.476 & $-$1.671 & &  \\
    &&                    &&          &        &    &&  3845.458 & $-$0.663 & &  \\

\noalign{\smallskip}
\hline
\end{tabular}
\end{table*}

\begin{table*}
\caption{Calculated hyperfine structure for the Cu~I 3247.53 and 3273.95\,\AA\ lines.}
\label{hfsCu2}
\centering
\begin{tabular}{cccccccccc}
\hline
\noalign{\smallskip}
\multicolumn{3}{c}{3247.53\AA;  $\chi$\,$=$\,0.00 eV} && \multicolumn{3}{c}{3273.95\AA; $\chi$\,$=$\,0.00 eV}   \\
\multicolumn{3}{c}{log gf(total) = $-$0.062} && \multicolumn{3}{c}{log gf(total) = $-$0.359} \\
\noalign{\smallskip}
\noalign{\smallskip}
\cline{1-3} \cline{5-7}  \\
$\lambda$ (\AA) & log gf & iso && $\lambda$ (\AA) & log gf & iso \\
\noalign{\smallskip} \\
\cline{1-3} \cline{5-7}  \\
 3247.507 &  $-$1.420 & 63 && 3273.925& $-$1.720 & 63 \\
 3247.506 &  $-$1.022 & 63 && 3273.923& $-$1.021 & 63 \\
 3247.505 &  $-$1.022 & 63 && 3273.966& $-$1.021 & 63 \\
 3247.547 &  $-$1.721 & 63 && 3273.965& $-$1.021 & 63 \\
 3247.546 &  $-$1.022 & 63 && 3273.923& $-$2.066 & 65 \\
 3247.544 &  $-$0.575 & 63 && 3273.922& $-$1.367 & 65 \\
 3247.506 &  $-$1.766 & 65 && 3273.968& $-$1.367 & 65 \\
 3247.505 &  $-$1.368 & 65 && 3273.966& $-$1.367 & 65 \\
 3247.503 &  $-$1.368 & 65 &&         &        &  \\
 3247.549 &  $-$2.067 & 65 &&         &        &  \\
 3247.547 &  $-$1.368 & 65 &&         &        &  \\
 3247.545 &  $-$0.921 & 65 &&         &        &  \\

\noalign{\smallskip}
\hline
\end{tabular}
\end{table*}

\section{Line fits to Arcturus and CS~31082-001}

In support of the analysis in Section~4, here we show the model fits to the 
reference star Arcturus for Co and Cu, and our fits to Sc, Ti, Mn, Co, Ni and Zn for CS~31082-001.
\FloatBarrier

\begin{figure}
    \centering
    \includegraphics[width=3.3in]{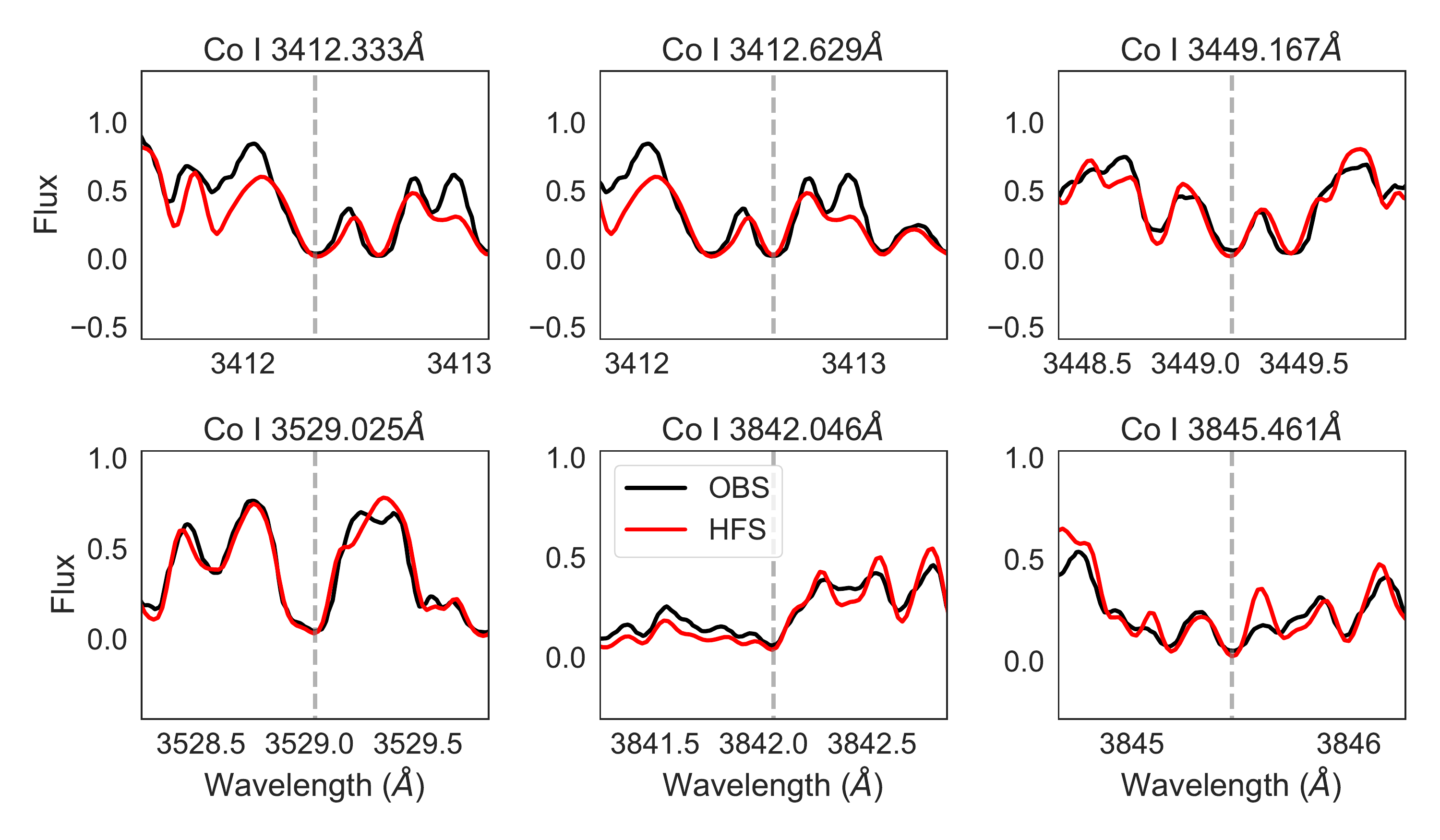}
    \caption{Acturus: Fits to the six selected lines of Co~I.}
    \label{ArcCo}

    \centering
    \includegraphics[width=3.3in]{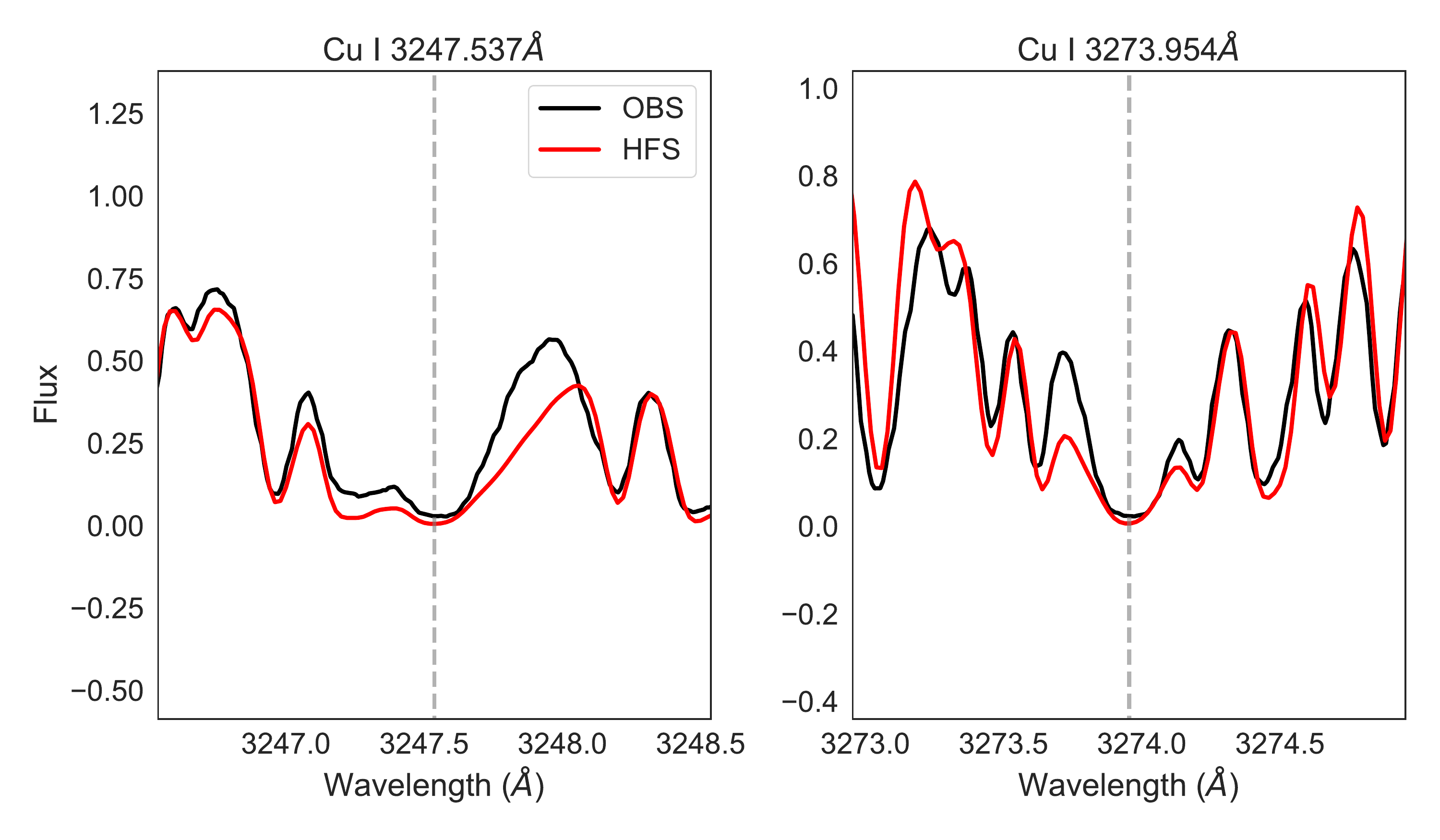}
    \caption{Arcturus: Fits to Cu~I 3247.537 and 3273.954 {\rm \AA} lines.}
    \label{ArcCu}
\end{figure}

\begin{figure}
    \centering
    \includegraphics[width=3.0in]{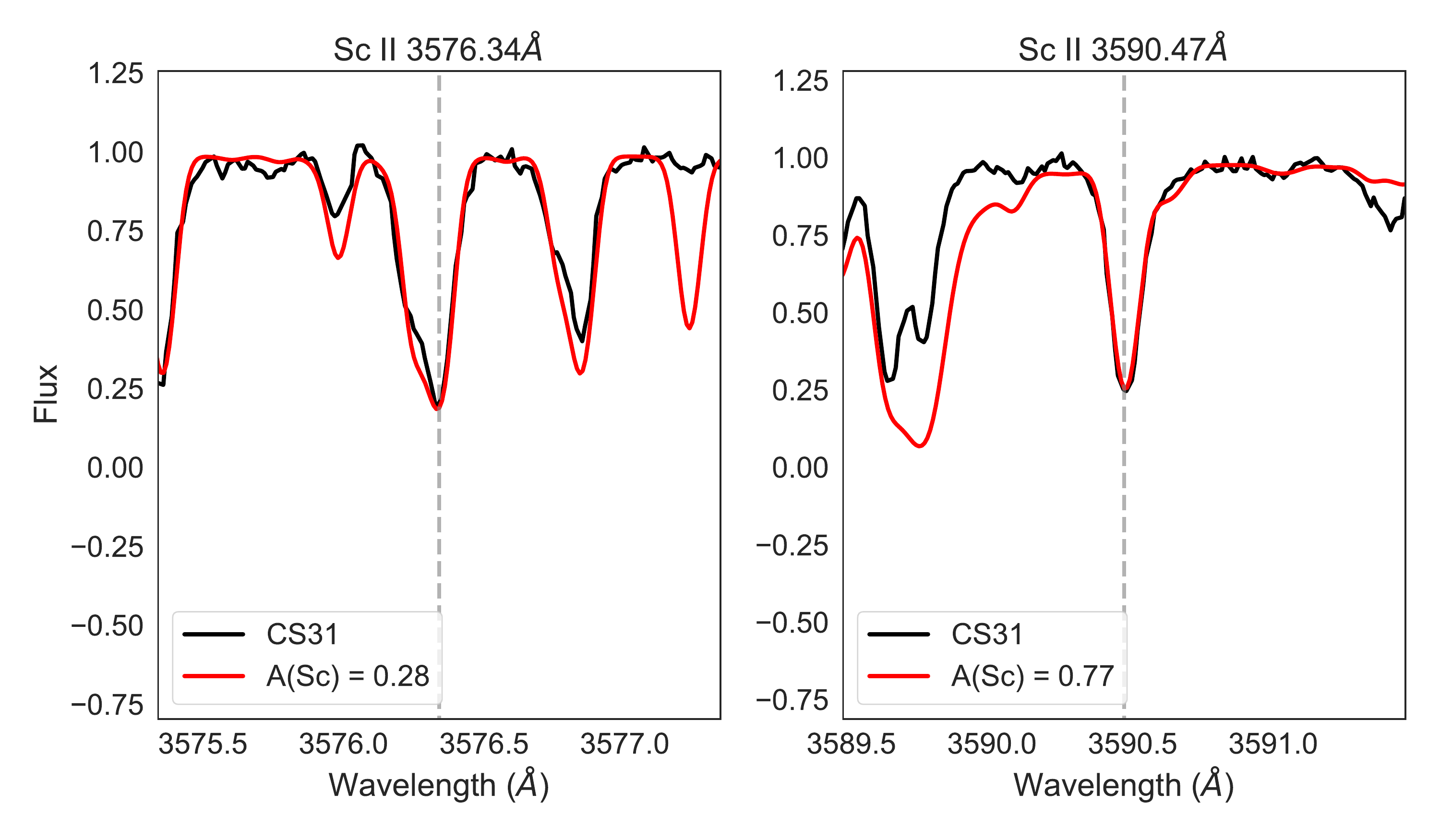}
    \caption{ Best-fit abundances for the Sc~II lines computed with [Sc/Fe]\,$=$\,+0.03 for the Sc~II 3576.34\,\AA\ line and [Sc/Fe]\,$=$\,+0.52 for the Sc~II 3590.47\,\AA\ line (without hyperfine structure).}
    \label{Sc}

    \centering
    \includegraphics[width=3.0in]{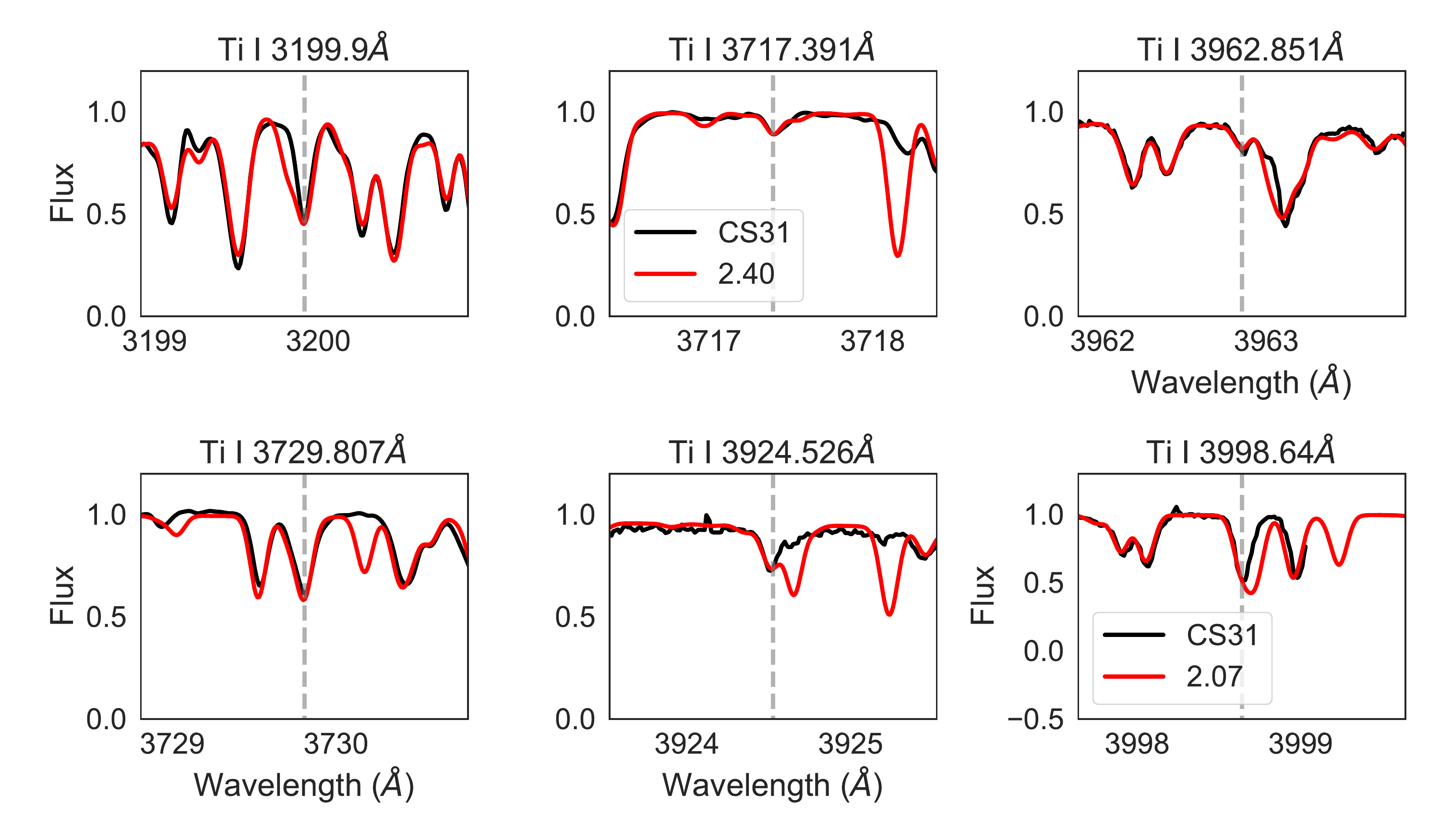}
    \caption{ Best-fit abundances for the Ti~I lines, computed with [Ti/Fe]\,$=$\,+0.35, whereas for the Ti~I 3998.64\,\AA\ line a best fit is found for [Ti/Fe]= +0.02.}
    \label{TiI}

    \centering
    \includegraphics[width=3.0in]{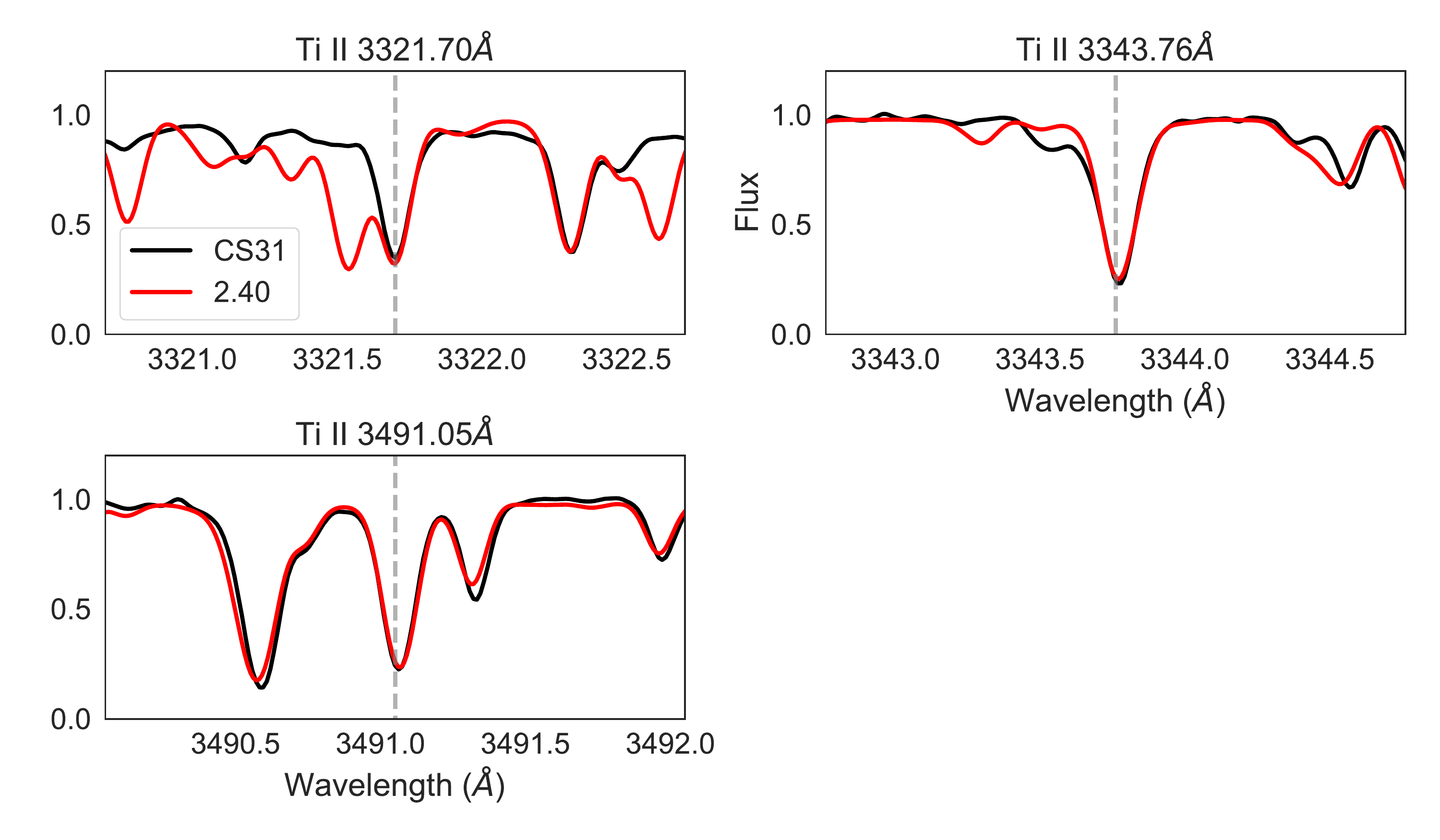}
    \caption{ Best-fit abundance for the Ti~II lines, computed with [Ti/Fe]\,$=\,$+0.35.}
    \label{Ti}

    \centering
    \includegraphics[width=3.0in]{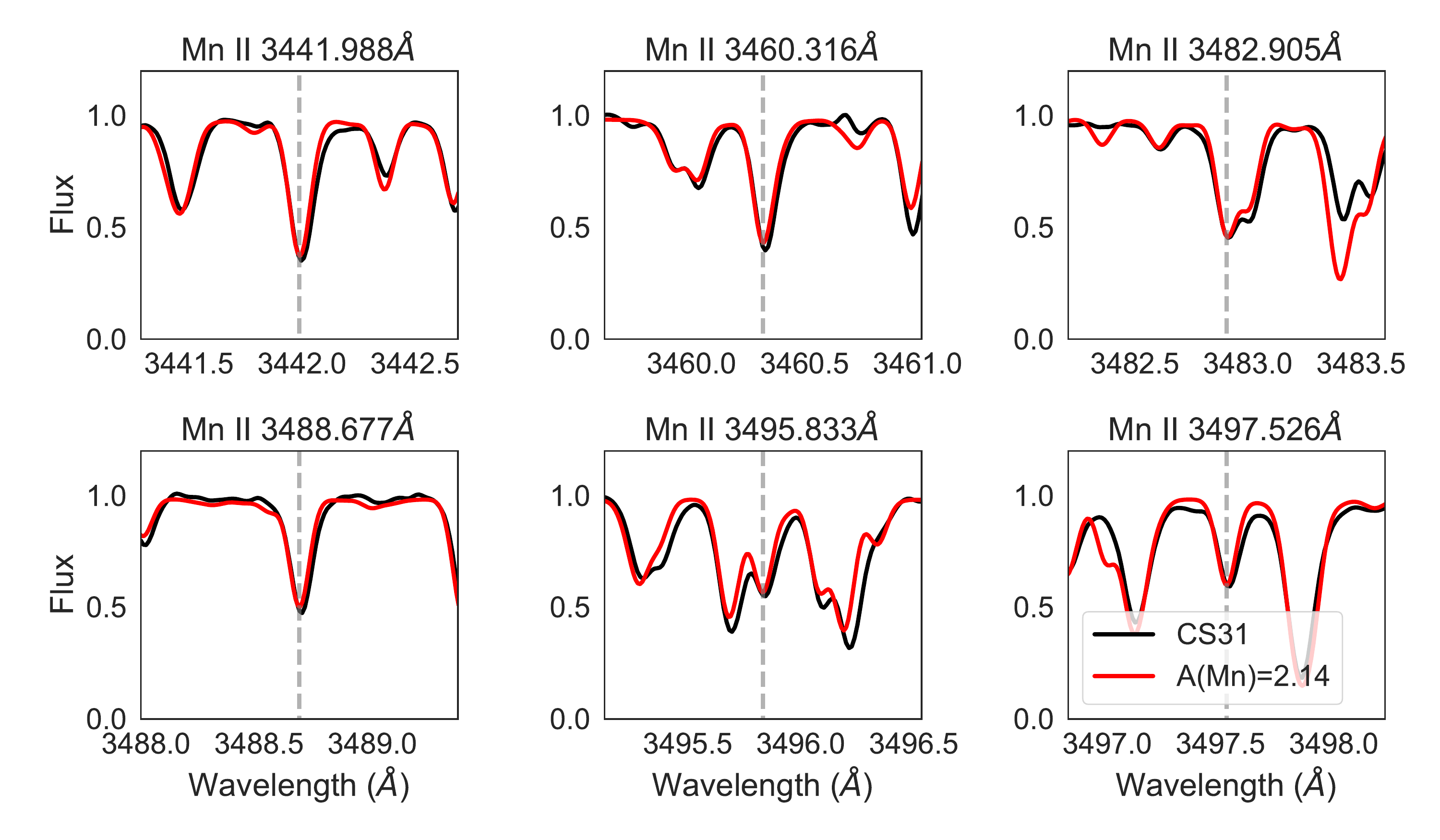}
    \caption{ Best-fit abundance for the Mn~II lines, computed with [Mn/Fe]\,$=$\,$-$0.39 (without hyperfine structure).}
    \label{Mn}
\end{figure}    

\begin{figure}
    \centering
    \includegraphics[width=3.3in]{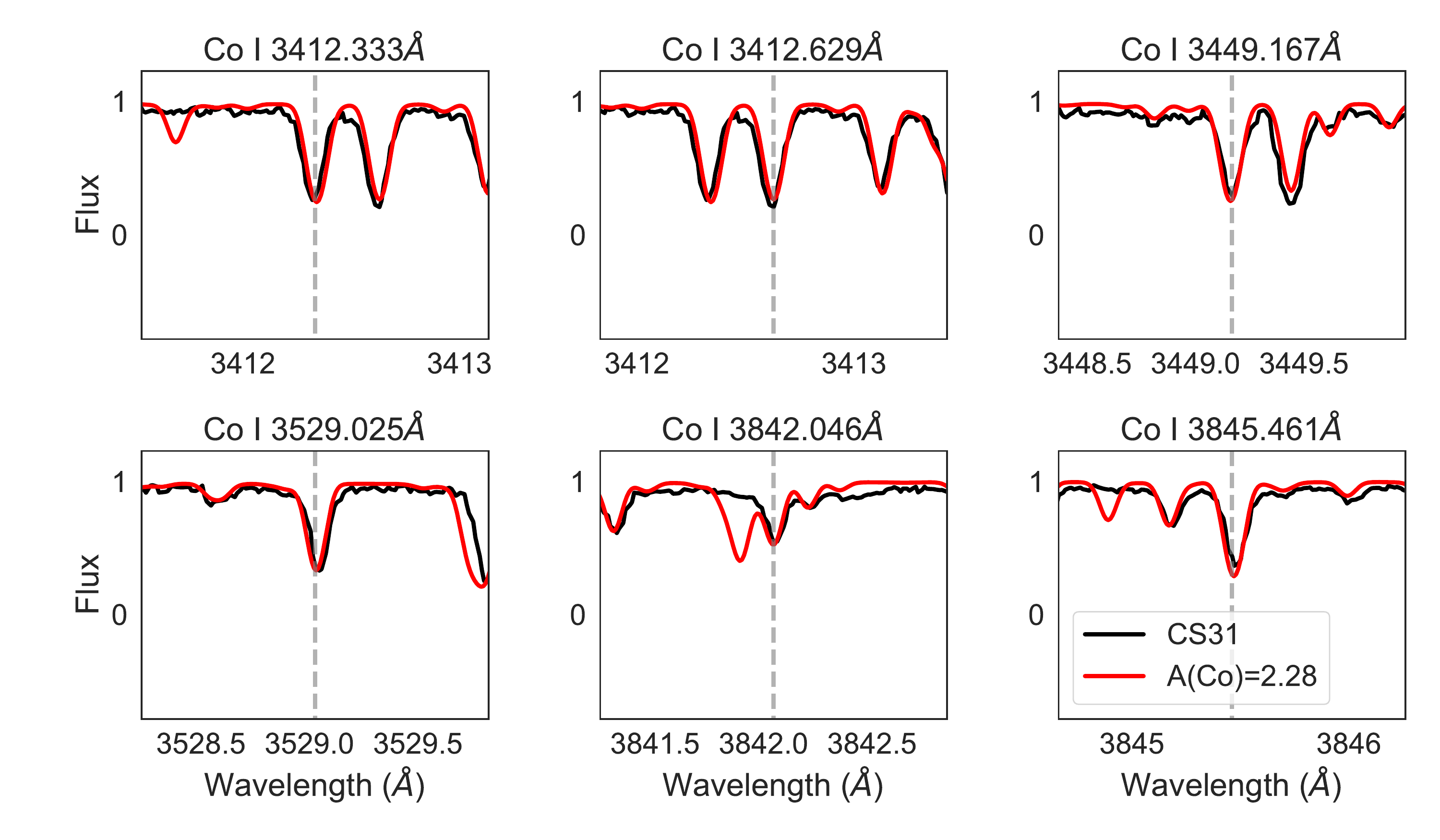}
    \caption{ Best-fit abundance for the Co~I lines, computed with [Co/Fe]\,$=$\,+0.19 (including hyperfine structure).}
    \label{Co}

    \centering
    \includegraphics[width=3.3in]{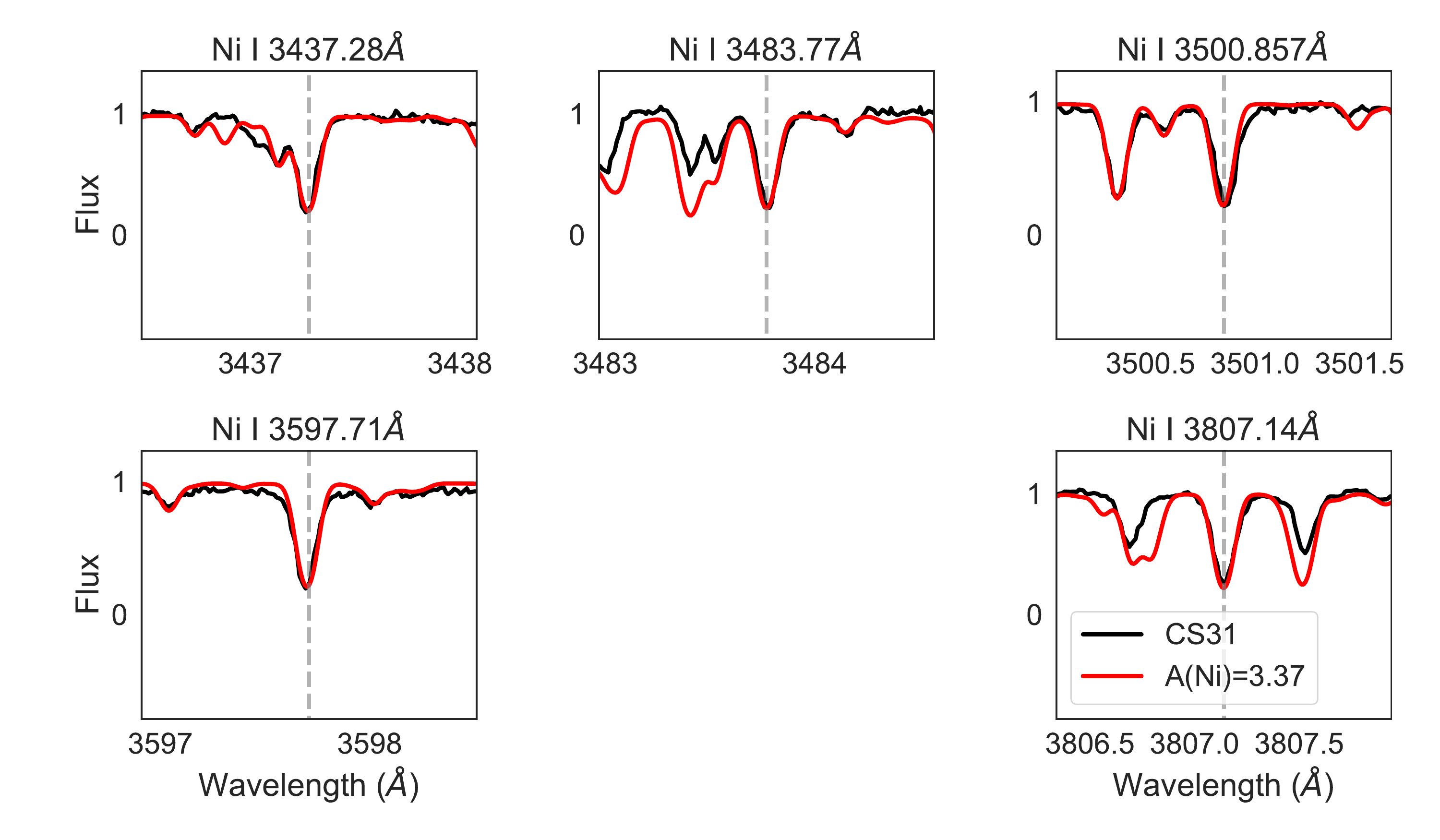}
    \caption{ Best-fit abundance for the Ni~I lines, computed with [Ni/Fe]\,$=$\,+0.05. }
    \label{Ni}

    \centering
    \includegraphics[width=3.3in]{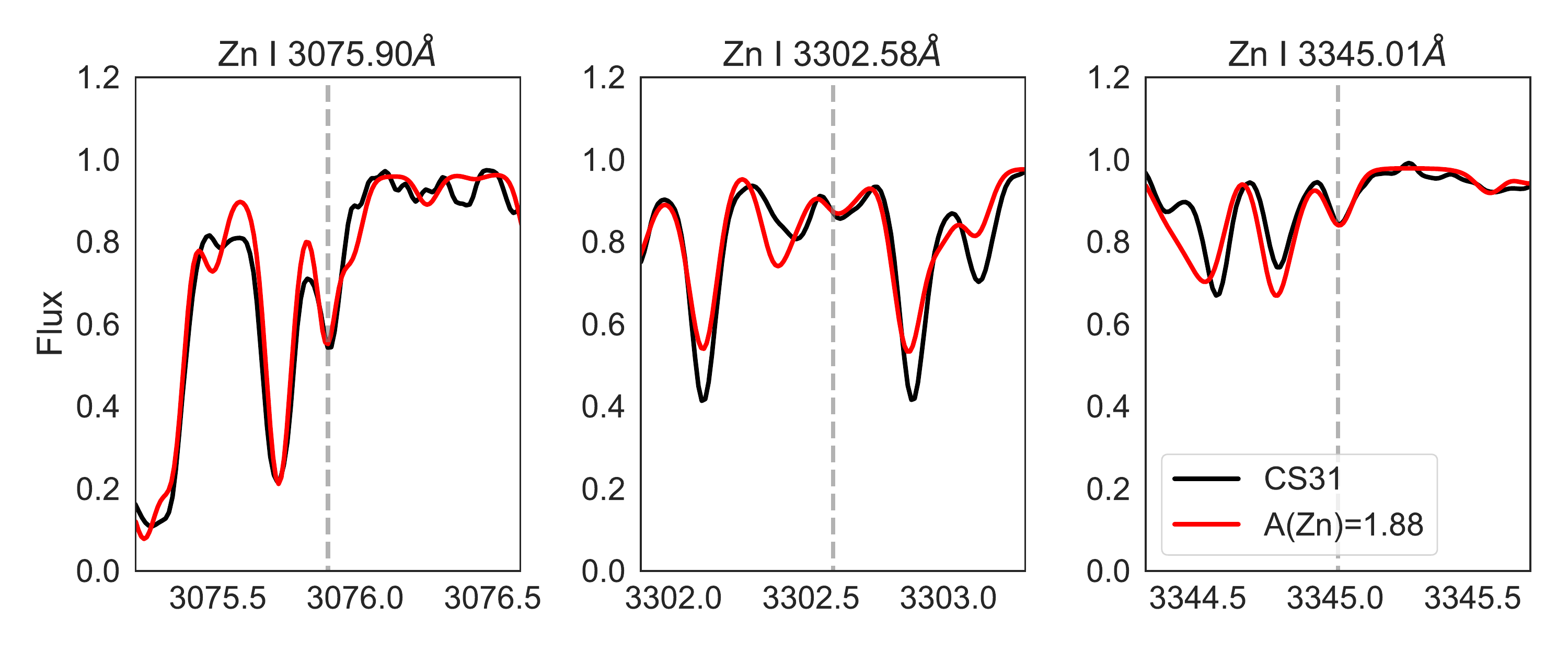}
    \caption{ Best-fit abundance for the Zn~I lines, computed with [Zn/Fe]\,$=$\,+0.22.}
    \label{Zn}
    \medskip
\end{figure}

\FloatBarrier
\section{Metal-poor stars with CUBES}
There are only a small number of extremely metal-poor stars that are sufficiently bright that we can attempt detailed abundance studies such as that presented here for CS~31082-001. The development of the new CUBES instrument for the VLT will offer significantly better sensitivity than current facilities at near-UV wavelengths, opening-up studies of a much larger sample of metal-poor stars for comparison with chemical-evolution models. To quantify this better we have used the spectrum of CS~31082-001 as a template to explore the future contribution of CUBES to this topic.

The construction of CUBES will start in early 2022, with science operations expected in 2028. It will be installed at the Cassegrain focus of one of the Unit Telescopes of the VLT, and the Phase A design of the instrument was outlined by Zanutta et al. (2022). In brief, a two-channel spectrograph will provide a continuous near-UV spectrum over the 300-405\,nm range, with the light split using a dichroic filter into blue and red arms (spanning 300-352\,nm and 346-405\,nm, respectively). An image slicer after the fore-optics will reduce the size of the optical image to enable the required high spectral resolution, which ranges from $R$\,$=$\,22,000 to 27,000 across the coverage of both the blue and red arms. The anticipated resolution at the wavelength of the Be doublet (313\,nm) is $R$\,$\sim$\,23,000, with a resolution element sampled by 2.3\,pixels on the detector. A fibre-link to the UVES instrument has also been studied to potentially allow simultaneous observations at longer visible wavelengths. As discussed by Zanutta et al. (2022), the optimised-design of CUBES for the maximum near-UV sensitivity should deliver at least a factor of ten better performance in end-to-end efficiency than UVES. 

As an example of the limits of observations with UVES, we highlight the recent study of BPS~BS~16968-0061 by Smiljanic et al. (2021). This very metal-poor star ([Fe/H]\,$\sim$\,$-$3.0) has $V$\,$=$\,13.22\,mag and required a total of 20\,hrs of UVES observations to achieve a S/N\,$\sim$\,200 per pixel in the region of the Be doublet. Using the exposure time calculator (ETC)\footnote{http://archives.ia2.inaf.it/cubes/\#/etc} developed alongside the CUBES design (see Genoni et al. 2022) and adopting a stellar template with parameters comparable to those of BPS~BS~16968-0061 (T$_{\rm eff}$\,$=$\,6500\,K, log\,$g$\,$=$\,4.0, [Fe/H]\,$=$\,$-$3.5) yields a S/N\,$\sim$\,200 in just 1.25\,hrs. While the exact parameters will undoubtedly evolve as the design progresses, this neatly illustrates the potentially large gain of CUBES at these short wavelengths. This means that quantitative analysis will be near trivial for stars with $V$\,$=$\,13\,mag that are only just feasible today with large, multi-night programmes; equally, S/N\,$=$\,200 will be possible for stars with $V$\,$=$\,15\,mag in one night of observations, and down to $V$\,$\simgreat$\,16\,mag for individual targets of note.

 To highlight the potential performance of CUBES for a fainter star similar to CS~31082-001, we investigated simulations using both model spectra and the observed UVES data.




\subsection{CUBES simulations with model spectra}\label{sims}

In our first tests we used the same synthetic spectra from {\tt Turbospectrum} used in the 
analysis in Section~4. By using synthetic models as inputs for instrument simulations, we are able to control all of the physical parameters (including abundances) and the noise characteristics of the template. We employed the CUBES end-to-end simulator (Genoni et al. 2022) to generate  simulated observations for different exposure times and magnitudes, with a goal of recovering stellar abundances for different species to within 0.1\,dex. 

A first qualitative study of near-UV diagnostic lines in this context was presented by Ernandes et al. (2020). A more detailed study of some of the heavy-element lines has recently been presented by Ernandes et al. (2022). Here we lighlight their results for Ge~I 3039\,\AA, which is observationally challenging given its proximity to the atmospheric cut-off and decline of the flux distribution of CS~31082-001 towards shorter wavelengths due to its low effective temperature. Nonetheless, Ernandes et al. (2022) concluded that Ge abundances (to $\pm$0.1\,dex) should be possible in a few hours of observations for a star that is two magnitudes fainter than CS~31082-001. For species at longer wavelengths (e.g. Hf), the equivalent gain in depth for targets that are now feasible within observations of a few hours increases to approx. three magnitudes, bringing many more targets within our reach than currently possible.

\subsection{Convolving the UVES data to the CUBES resolution}\label{conv}

\begin{figure}
    \centering
    \includegraphics[width=3.3in]{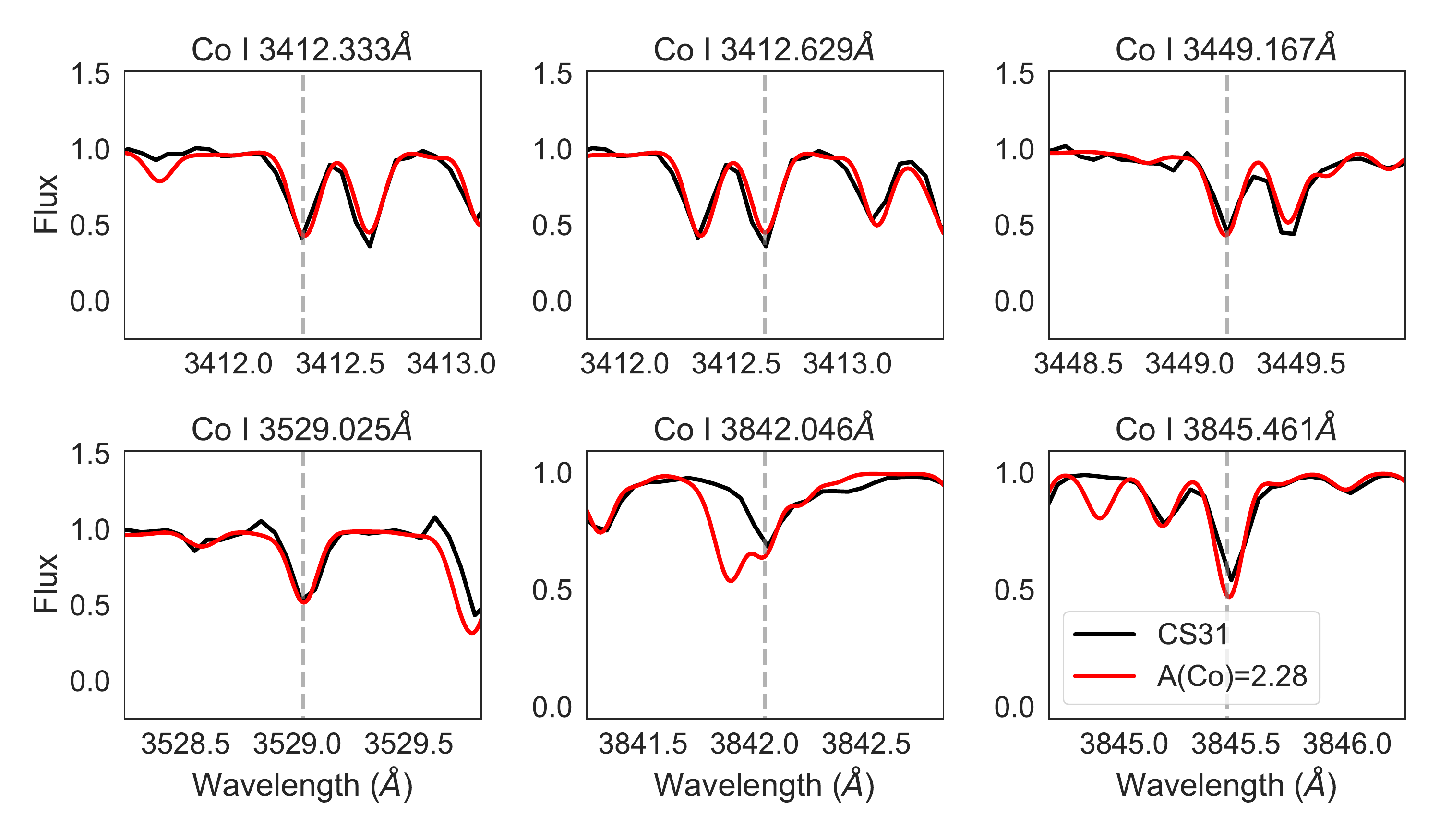}
    \includegraphics[width=3.3in]{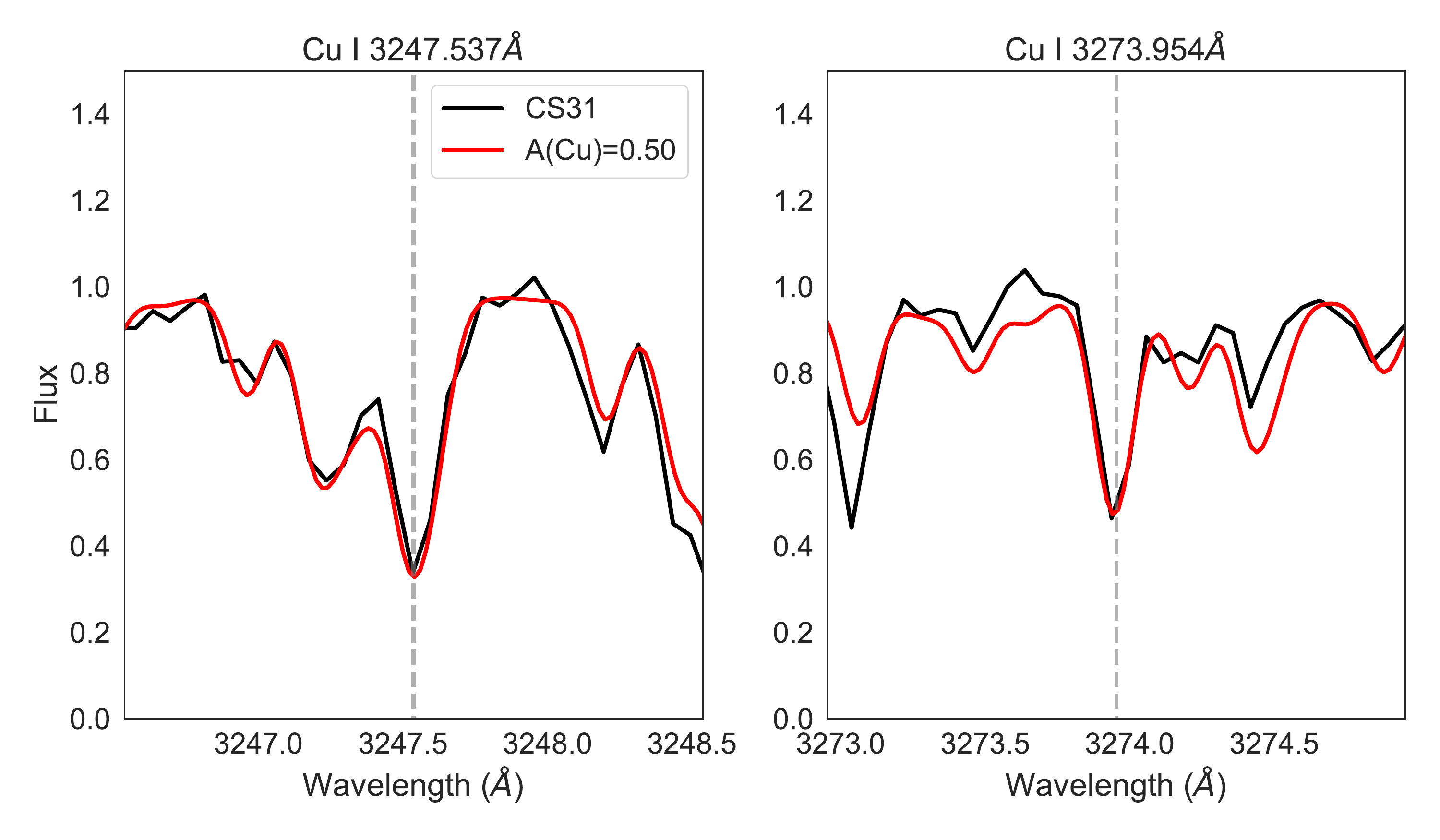}
    \caption{Line fits for the Co and Cu lines from the UVES spectrum degraded to the resolution of CUBES. }
    \label{CUBESCOCU}
\end{figure}

Another approach is to degrade the resolution of the UVES spectrum to that of CUBES, i.e. from $R$\,$\sim$\,77,000 to $\sim$\,24,000. Of course, smoothing and then rebinning to a larger pixel scale results in an increase of the S/N per pixel compared to the UVES spectrum. We then added artificial noise to the degraded spectrum to match the S/N of the original UVES data.
For most of the lines analysed in the current study, we were able to recover the same abundance as obtained with UVES. The exception is the Be line, where the weaker line in the doublet becomes too blended. Some examples are given in Fig.~\ref{CUBESCOCU}, with the same S/N per pixel as the UVES spectrum.

The error can be estimated using the formula from Cayrel (1988), in its final form as given in by Cayrel et al. (2004): $\sigma={\frac{1.5}{S/N}} \sqrt{FWHM*\delta{x}}$, where $\delta{x}$ is the pixel size. For $R$\,$=$\,24,000 we have a FWHM\,$=$\,0.1429 {\rm \AA} and $\delta{x}$\,$=$\,0.06 {\rm \AA} in the spectra. Assuming a mean
S/N=100, 
we get an error of $\Delta$EW $\sim$ 1.4 m{\rm \AA}. Given that this formula neglects the uncertainty in continuum placement, as well as the uncertainty in the determination of the FWHM of the lines, this argues for $\Delta$EW $\sim$ 3.0 to 4.5 m{\rm \AA} for CUBES observations with S/N$\sim$100. As an example, the EW of the Be~II 3131 {\rm \AA} line is 3 to 5 mA, giving A(Be)\,$=$\,$-$2.5 to $-$2.8, i.e. the error is the same order of magnitude as the EW itself. Assuming the result of A(Be)\,$=$\,$-$2.5 for CS 31082-001, the estimate using CUBES would be in the range A(Be)\,$=$\,$-$2.2 to $-$3.0.

\end{document}